# Performance of Multilevel Flash Memories with Different Binary Labelings: A Multi-User Perspective

Pengfei Huang, *Student Member, IEEE,* Paul H. Siegel, *Fellow, IEEE,* and
Eitan Yaakobi, *Member, IEEE*

### Abstract

In this work, we study the performance of different decoding schemes for multilevel flash memories where each page in every block is encoded independently. We focus on the multi-level cell (MLC) flash memory, which is modeled as a two-user multiple access channel suffering from asymmetric noise. The uniform rate regions and sum rates of Treating Interference as Noise (TIN) decoding and Successive Cancelation (SC) decoding are investigated for a Program/Erase (P/E) cycling model and a data retention model. We examine the effect of different binary labelings of the cell levels, as well as the impact of further quantization of the memory output (i.e., additional read thresholds). Finally, we extend our analysis to the three-level cell (TLC) flash memory.

## I. Introduction

NAND flash memory is a promising non-volatile data storage medium, and has been widely used in customer electronics as well as enterprise data centers. It has many advantages over traditional magnetic recording, e.g., higher read throughput and less power consumption [3]. The basic storage unit in a NAND flash memory is a floating-gate transistor referred to as a cell. The voltage levels of a cell can be adjusted by a program operation and are used to represent the stored data. The cells typically have 2, 4, and 8 voltage levels (1, 2, and 3 bits/cell respectively) and are referred to as single-level cell (SLC), multi-level cell (MLC), and three-level cell (TLC) respectively. Cells are grouped into pages, which are grouped into blocks. A page is the smallest unit for read and write operations, while a block is the smallest unit for an erase operation.

P. Huang and P.H. Siegel are with the Department of Electrical and Computer Engineering and the Center for Memory Recording Research, University of California, San Diego, La Jolla, CA 92093, U.S.A. (e-mail: {pehuang, psiegel}@ucsd.edu).

E. Yaakobi is with the Department of Computer Science, Technion – Israel Institute of Technology, Haifa 32000, Israel (e-mail: yaakobi@cs.technion.ac.il).

This research was supported by the CMRR at UCSD and NSF Grants CCF-1116739 and CCF-1405119. Part of the results in the paper will appear at the IEEE ISIT, Barcelona, Spain, July 10-15, 2016 (reference [10]).

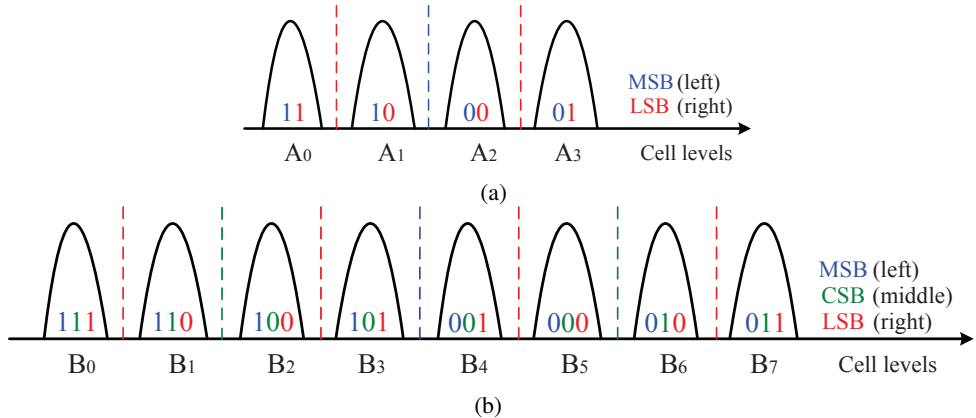

Fig. 1. (a) The four voltage levels and Gray labeling for a cell in MLC flash memories. A total of three reads are employed for decoding two pages. (b) The eight voltage levels and Gray labeling for a cell in TLC flash memories. A total of seven reads are employed for decoding three pages.

The two bits belonging to a multi-level cell (MLC) are separately mapped to two pages. The most significant bit (MSB) is mapped to the lower page while the least significant bit (LSB) is mapped to the upper page. As shown in Fig. 1(a), we represent the four voltage levels in MLC flash memory as $A_0$, $A_1$, $A_2$, and $A_3$ in increasing order of voltage levels. The corresponding 2-bit patterns written to the lower page (MSB) and upper page (LSB) are '11', '10', '00', and '01', which are called *Gray* labeling. Similarly, the three bits belonging to a three-level cell (TLC) are separately mapped to three pages. We refer to the first bit as the most significant bit (MSB), the second bit as the center significant bit (CSB), and the third bit as the least significant bit (LSB). As shown in Fig. 1(b), we represent the eight voltage levels in TLC flash memory as $B_0$, $B_1$, $B_2$, $B_3$, $B_4$, $B_5$, $B_6$, and $B_7$ in increasing order of voltage levels. The corresponding 3-bit patterns for the MSB, CSB, and LSB also use Gray labeling and are '111', '110', '100', '101', '001', '000', '010', and '011', respectively.

The channel characterization of flash memory is important for understanding the fundamental density limits as well as designing effective signal processing algorithms and error-correcting codes (ECC) [2], [7], [8], [16]. Many experiments have shown that the distribution of a readback signal (i.e., the voltage level of a cell) in flash memories is asymmetric [4], [5], [13], [19]. In [14], a mixed normal-Laplace distribution model was proposed and validated to capture this asymmetry feature well. Based on a certain type of voltage level distribution, in [11], [12], [18], [20], [21], the capacity of a flash memory was studied by treating each cell as an input variable. For example, the capacity of MLC flash memory was analyzed in [21] by modeling MLC flash memory as a 4-ary input point-to-point channel with additive white Gaussian noise. Moreover, in the same paper, multiple reads were used to obtain soft information, thus improving the decoding performance. Similarly, the capacity of TLC flash memory was recently studied in [18] by considering TLC flash memory as an 8-ary input point-to-point channel with asymmetric mixed normal-Laplace noise.

However, in current MLC (or TLC) flash memories, 2 (or 3) bits in a cell are mapped to 2 (or 3) pages, which are actually encoded independently. Hence, previous works [11], [12], [18], [20], [21] based on a point-to-point channel model only give an upper bound on the sum rate of all pages. In this paper, we take a different perspective, and model the flash memory as a multi-user system, where each user corresponds to a page and is encoded independently. To the best of our knowledge, this is the first time that flash memories have been studied in this framework.

Our goal is to study the fundamental performance limits of flash memories with different decoding schemes. Here, we consider both low-complexity Treating Interference as Noise (TIN) decoding and relatively high-complexity Successive Cancelation (SC) decoding, and derive the conditions when the sum rate of TIN decoding equals that of SC decoding. Then, achievable rate regions and sum rates of both decoding schemes are determined for different channel models, represented by channel transition matrices from cell voltage levels to quantized readback outputs. The effect of different binary labelings of the cell levels is also studied, and the optimal labeling for each decoding scheme and channel model is identified. It is shown that TIN and SC decodings both outperform the Default Setting (DS) decoding, a model of current flash memory technology, which uses Gray labeling of cell levels, along with separate quantization and decoding of each page. Moreover, the impact of further quantization of the memory output (i.e., additional read thresholds) is analyzed, and the performance improvement is validated by various simulations. Although in this paper we focus on information-theoretical analysis of flash memories, some of our results give us an insight on very simple ECC solutions, which are attractive for practical applications.

The remainder of the paper is organized as follows. In Section II, we model the multilevel flash memory as a multi-user system. In Section III, we introduce three different decoding schemes for MLC flash memories. In Section IV, we study the decoding performance of MLC flash memories for different channel models. Different labelings and multiple reads are discussed. In Section V, we further investigate TLC flash memories. We conclude the paper in Section VI.

Throughout the paper, we follow the notations in [9]. Random variables are denoted with upper case letters (e.g., $X$) and their realizations with lower case (e.g., $x$). Calligraphic letters (e.g., $\mathcal{X}$) are used for finite sets. $\mathbb{R}$ is the real line. The discrete interval $[i : j]$ is defined as the set $\{i, i+1, \ldots, j\}$. For an $n$-length vector $v$, $v(i)$ represents the value of its $i$th coordinate, $i = 1, 2, \ldots, n$. We use the notation $p(x)$ to abbreviate the probability $P(X = x)$, and likewise for conditional and joint probabilities of both scalar and vector random variables, e.g., $p(y|x) = P(Y = y|X = x)$. For a probability vector $(\frac{p_1}{p_s}, \frac{p_2}{p_s}, \ldots, \frac{p_n}{p_s})$ where $p_s = \sum_{i=1}^n p_i$ and $p_i \geqslant 0$, $i = 1, 2, \ldots, n$, the entropy function $H(\frac{p_1}{p_s}, \frac{p_2}{p_s}, \ldots, \frac{p_n}{p_s}) = -\sum_{i=1}^n \frac{p_i}{p_s} \log_2 \frac{p_i}{p_s}$. We will also use the function $f(x) = x \log_2 x$. Therefore, we can express $H(\frac{p_1}{p_s}, \frac{p_2}{p_s}, \ldots, \frac{p_n}{p_s}) = -\frac{1}{p_s} \sum_{i=1}^n f(p_i) + \log_2 p_s$. For other notations in this paper, they will be stated before use.

## II. System Model for Multilevel Flash Memories

We model a multilevel flash memory as a $k$-user multiple access channel with $k$ independent inputs $X_1, \cdots, X_k$, and one output $Y$ ($k = 2$ for MLC flash, and $k = 3$ for TLC flash).

Specifically, the readback signal $\tilde{Y} \in \mathbb{R}$ in a flash memory is expressed as

$$\tilde{Y} = \sigma(X_1, \cdots, X_k) + Z, \tag{1}$$

where $X_1, \cdots, X_k \in \{0, 1\}$ represent data from $k$ independent pages, $Z \in \mathbb{R}$ stands for the asymmetric noise (see [14] for more details on the normal-Laplace distribution model), and $\sigma$ maps an input $(x_1, \cdots, x_k)$ to a voltage level $v$. More specifically, $\sigma$ is a bijective mapping from the set $\mathcal{T}$ which consists of all $k$-length binary strings to the set $\mathcal{V}$ which consists of $2^k$ voltage level values. For $k = 2$ (MLC flash), $\mathcal{T}_{MLC} = \{11, 10, 01, 00\}$ and $\mathcal{V}_{MLC} = \{A_0, A_1, A_2, A_3\}$. By a slight abuse of notation, we write the mapping $\sigma$ as a vector $\sigma = (w_0, w_1, w_2, w_3)$ (where the $w_i$, $i = 0, 1, 2, 3$, represent the full set of possible 2-tuples) to represent the mapping $\sigma(w_i) = A_i$ for $i = 0, 1, 2, 3$. For example, $\sigma = (11, 10, 00, 01)$ means $\sigma(11) = A_0$, $\sigma(10) = A_1$, $\sigma(00) = A_2$, and $\sigma(01) = A_3$. Similarly, for $k = 3$ (TLC flash), $\mathcal{T}_{TLC} = \{111, 110, 101, 100, 011, 010, 001, 000\}$ and $\mathcal{V}_{TLC} = \{B_0, B_1, \ldots, B_7\}$. We write $\sigma = (w_0, w_1, \ldots, w_7)$ (where the $w_i$, $i = 0, 1, \ldots, 7$, represent the full set of possible 3-tuples) to represent the mapping $\sigma(w_i) = B_i$ for $i = 0, 1, \ldots, 7$.

During the readback process, a quantizer $Q$ is used to quantize $\tilde{Y}$ to obtain an output $Y$, i.e., $Y = Q(\tilde{Y})$, where the function $Q(\cdot)$ is a mapping from $\mathbb{R}$ to a finite alphabet set $\mathcal{Y} = \{s_0, s_1, \ldots, s_{q-1}\}$ of cardinality $q$. Usually $q = 2^k$, but this is not necessary. The cardinality $q$ can correspond to a large number by applying multiple reads. From an information-theoretical point of view, this means that more soft information is obtained for decoding.

In this paper, we refer to a *mapping $\sigma$* as a *labeling*.

## III. Decoding Schemes for MLC Flash Memories

In this section, we investigate three decoding schemes for MLC flash memories.

Given a labeling $\sigma$ and a quantizer $Q$, the MLC flash memory channel can be modeled as a 2-user discrete memoryless multiple access channel $\mathcal{W}_{MLC}$: $(\mathcal{X} \times \mathcal{X}, p(y|x_1, x_2), \mathcal{Y})$, where $\mathcal{X} = \{0, 1\}$, $\mathcal{Y} = \{s_0, s_1, \ldots, s_{q-1}\}$, and $p(y|x_1, x_2)$ is the transition probability for any $x_1, x_2 \in \mathcal{X}$ and $y \in \mathcal{Y}$. For simplicity, define conditional probability $P(Y = y | X_1 = x_1, X_2 = x_2) \overset{\text{def}}{=} p_{BD(x_1,x_2),y}$, where $BD(\cdot)$ is a function that converts a binary string into its decimal value, e.g., $P(Y = s_0 | X_1 = 1, X_2 = 0) = p_{2,s_0}$.

Users $j = 1, 2$ independently encode their messages $M_j$ into corresponding $n$-length codewords $x_j^n$ and store them over the shared channel (i.e., a set of cells) for the reader (receiver). Following the notation in [9], we define a $(2^{nR_1}, 2^{nR_2}, n)$ code by two encoders $x_1^n(m_1)$ and $x_2^n(m_2)$ for messages $m_1$ and $m_2$ from message sets $[1 : 2^{nR_1}]$ and $[1 : 2^{nR_2}]$ respectively, and a decoder that assigns an estimate $(\hat{m}_1, \hat{m}_2)$ based on the received sequence $y^n$. We assume that the message pair $(M_1, M_2)$ is uniform over $[1 : 2^{nR_1}] \times [1 : 2^{nR_2}]$. The average probability of error is defined as $\mathrm{P}_e^{(n)} = P\{(M_1, M_2) \neq (\hat{M}_1, \hat{M}_2)\}$. A rate pair $(R_1, R_2)$ is said to be achievable if there exists a sequence of $(2^{nR_1}, 2^{nR_2}, n)$ codes such that $\lim_{n \to \infty} \mathrm{P}_e^{(n)} = 0$. The capacity region is the closure of the set of achievable rate pairs $(R_1, R_2)$.

The capacity region of this multiple access channel is fully characterized [6], [9]. However, in this paper, to make the analysis simple and yet representative, we are interested in the *uniform rate region* for different decoding schemes. For a multiple access channel, the uniform rate region is the achievable region corresponding to the case that the input distributions are uniform. For other input distributions, the analysis is similar.

For a channel $\mathcal{W}_{MLC}$, the **Treating Interference as Noise (TIN)** decoding scheme decodes $X_1$ and $X_2$ independently based on $Y$ [6], [9]. Its uniform rate region $\mathscr{R}^{TIN}$ for lower page $X_1$ and upper page $X_2$ is the set of all pairs $(R_1, R_2)$ such that 1) $0 \leqslant R_1 \leqslant I(X_1; Y)$ and 2) $0 \leqslant R_2 \leqslant I(X_2; Y)$. In $\mathscr{R}^{TIN}$, the sum rate[1] $r_s^{TIN} = \max\{R_1 + R_2 : (R_1, R_2) \in \mathscr{R}^{TIN}\} = I(X_1; Y) + I(X_2; Y)$.

For a channel $\mathcal{W}_{MLC}$, the **Successive Cancelation (SC)** decoding scheme decodes $X_1$ and $X_2$ in some order based on $Y$ [6], [9]. Its uniform rate region $\mathscr{R}^{SC}$ for lower page $X_1$ and upper page $X_2$ is the set of all pairs $(R_1, R_2)$ such that 1) $R_1 \leqslant I(X_1; Y | X_2)$, 2) $R_2 \leqslant I(X_2; Y | X_1)$, and 3) $R_1 + R_2 \leqslant I(X_1, X_2; Y)$. In $\mathscr{R}^{SC}$, the sum rate $r_s^{SC} = \max\{R_1 + R_2 : (R_1, R_2) \in \mathscr{R}^{SC}\} = I(X_1, X_2; Y)$.

**Remark 1** For TIN decoding, $X_1$ and $X_2$ are decoded independently and can be implemented in parallel. However, for SC decoding, $X_1$ and $X_2$ are decoded in a certain order. In general, TIN decoding is preferred for its low decoding complexity, but the uniform rate region $\mathscr{R}^{TIN} \subseteq \mathscr{R}^{SC}$ and the sum rate $r_s^{TIN} \leqslant r_s^{SC}$. □

---

[1] In this paper, for the sake of brevity, we use the term "sum rate" to represent the maximum sum rate in the corresponding rate region.

The following theorem gives the condition when the sum rates of TIN decoding and SC decoding are the same.

**Theorem 1.** *For a channel $\mathcal{W}_{MLC}$, the sum rates satisfy $r_s^{TIN} \leqslant r_s^{SC}$ with equality if and only if $p_{3,s_j} p_{0,s_j} = p_{2,s_j} p_{1,s_j}$ for all $j = 0, 1, \ldots, q-1$. If $r_s^{TIN} = r_s^{SC}$, then $\mathscr{R}^{TIN} = \mathscr{R}^{SC}$ and the rate region is a rectangle.*

*Proof:* We bound the value $r_s^{SC} - r_s^{TIN}$ as follows

$$r_s^{SC} - r_s^{TIN} = I(X_1, X_2; Y) - I(X_1; Y) - I(X_2; Y) = I(X_2; Y|X_1) - I(X_2; Y)$$
$$= H(X_2|X_1) - H(X_2|X_1, Y) - \Big( H(X_2) - H(X_2|Y) \Big)$$
$$\overset{(a)}{=} I(X_1; X_2|Y) = \sum_{j=0}^{q-1} \Big( \sum_{i=0}^{3} \frac{p_{i,s_j}}{4} \Big) I(X_1; X_2|Y = s_j) \geqslant 0,$$

where in step $(a)$ we use $H(X_2|X_1) = H(X_2)$ which follows from the fact that $X_1$ and $X_2$ are independent.

Now, $I(X_1; X_2|Y) \geqslant 0$ with equality if and only if $X_1$ and $X_2$ are conditionally independent given $Y = s_j$, i.e., $P(X_1, X_2|Y = s_j) = P(X_1|Y = s_j)P(X_2|Y = s_j)$. Thus, we need to check four cases:

1) $P(X_1 = 1, X_2 = 1|Y = s_j) = P(X_1 = 1|Y = s_j)P(X_2 = 1|Y = s_j)$, i.e., $\frac{p_{3,s_j}}{\sum_{i=0}^{3} p_{i,s_j}} = \frac{p_{3,s_j} + p_{2,s_j}}{\sum_{i=0}^{3} p_{i,s_j}} \frac{p_{3,s_j} + p_{1,s_j}}{\sum_{i=0}^{3} p_{i,s_j}}$.

2) $P(X_1 = 1, X_2 = 0|Y = s_j) = P(X_1 = 1|Y = s_j)P(X_2 = 0|Y = s_j)$, i.e., $\frac{p_{2,s_j}}{\sum_{i=0}^{3} p_{i,s_j}} = \frac{p_{3,s_j} + p_{2,s_j}}{\sum_{i=0}^{3} p_{i,s_j}} \frac{p_{2,s_j} + p_{0,s_j}}{\sum_{i=0}^{3} p_{i,s_j}}$.

3) $P(X_1 = 0, X_2 = 1|Y = s_j) = P(X_1 = 0|Y = s_j)P(X_2 = 1|Y = s_j)$, i.e., $\frac{p_{1,s_j}}{\sum_{i=0}^{3} p_{i,s_j}} = \frac{p_{1,s_j} + p_{0,s_j}}{\sum_{i=0}^{3} p_{i,s_j}} \frac{p_{3,s_j} + p_{1,s_j}}{\sum_{i=0}^{3} p_{i,s_j}}$.

4) $P(X_1 = 0, X_2 = 0|Y = s_j) = P(X_1 = 0|Y = s_j)P(X_2 = 0|Y = s_j)$, i.e., $\frac{p_{0,s_j}}{\sum_{i=0}^{3} p_{i,s_j}} = \frac{p_{1,s_j} + p_{0,s_j}}{\sum_{i=0}^{3} p_{i,s_j}} \frac{p_{2,s_j} + p_{0,s_j}}{\sum_{i=0}^{3} p_{i,s_j}}$.

To satisfy conditions 1) – 4), we have $p_{3,s_j} p_{0,s_j} = p_{2,s_j} p_{1,s_j}$ for all $j = 0, 1, \ldots, q-1$.

Finally, assuming $r_s^{TIN} = r_s^{SC}$, i.e., $I(X_1; Y) + I(X_2; Y) = I(X_1, X_2; Y)$, since $I(X_1, X_2; Y) = I(X_1; Y) + I(X_2; Y|X_1) = I(X_2; Y) + I(X_1; Y|X_2)$, we have $I(X_1; Y) = I(X_1; Y|X_2)$ and $I(X_2; Y) = I(X_2; Y|X_1)$, which means $\mathscr{R}^{TIN} = \mathscr{R}^{SC}$. $\blacksquare$

The upper bound on the difference between $r_s^{SC}$ and $r_s^{TIN}$ is given by the following theorem.

**Theorem 2.** *For a channel $\mathcal{W}_{MLC}$, the rate difference $r_s^{SC} - r_s^{TIN} \leqslant 1$ with equality if and only if $p_{3,s_j} + p_{2,s_j} = p_{1,s_j} + p_{0,s_j}$ and $p_{3,s_j} p_{1,s_j} = p_{2,s_j} p_{0,s_j} = 0$ for all $j = 0, 1, \ldots, q-1$.*

*Proof:* We bound $r_s^{SC} - r_s^{TIN}$ as

$$r_s^{SC} - r_s^{TIN} = I(X_1, X_2; Y) - I(X_1; Y) - I(X_2; Y) = I(X_1; X_2|Y) = H(X_1|Y) - H(X_1|X_2, Y)$$
$$\overset{(a)}{\leqslant} H(X_1) - H(X_1|X_2, Y) \overset{(b)}{\leqslant} H(X_1) = 1,$$

where step (a) follows from $H(X_1|Y) \leqslant H(X_1)$, and step (b) is due to $H(X_1|X_2, Y) \geqslant 0$. Thus, $r_s^{SC} - r_s^{TIN} = 1$ if and only if 1) $H(X_1|Y) = H(X_1) = 1$, and 2) $H(X_1|X_2, Y) = 0$.

| $V$ | Inputs: $(X_1, X_2)$ | | | Output: $Y$ | | | |
|---|---|---|---|---|---|---|---|
| Levels | Gray | NO | EO | $s_0$ | $s_1$ | $s_2$ | $s_3$ |
| $A_0$ | (11) | (11) | (11) | $a_1$ | $1 - a_1$ | 0 | 0 |
| $A_1$ | (10) | (10) | (00) | 0 | $b_1$ | $1 - b_1$ | 0 |
| $A_2$ | (00) | (01) | (01) | 0 | 0 | $c_1$ | $1 - c_1$ |
| $A_3$ | (01) | (00) | (10) | 0 | 0 | 0 | 1 |

The condition $H(X_1|Y) = \sum_{j=0}^{q-1} \left( \sum_{i=0}^{3} \frac{p_{i,s_j}}{4} \right) H(X_1|Y = s_j) = 1$ holds if and only if $p_{3,s_j} + p_{2,s_j} = p_{1,s_j} + p_{0,s_j}$ for all $j = 0, 1, \ldots, q - 1$. Similarly, $H(X_1|X_2, Y) = 0$ requires that $p_{3,s_j} p_{1,s_j} = p_{2,s_j} p_{0,s_j} = 0$ for all $j = 0, 1, \ldots, q - 1$. ∎

The third decoding scheme we consider is modeled upon current MLC flash memory technology. For this scheme, the Gray labeling $\sigma = (11, 10, 00, 01)$ is used to map binary inputs $(X_1, X_2)$ to cell levels $V$. The lower page $X_1$ and upper page $X_2$ are decoded independently according to different quantizations and a total of three reads are employed: 1) to decode $X_1$, $\tilde{Y}$ is quantized by one read between voltage levels $A_1$ and $A_2$ (see Fig. 1(a)), and its corresponding output is $Y_1$; 2) to decode $X_2$, $\tilde{Y}$ is quantized by two reads between voltage levels $A_0$ and $A_1$, and between $A_2$ and $A_3$, respectively (see Fig. 1(a)), and its corresponding output is $Y_2$. We call this **Default Setting (DS)** decoding, and it is used as our *baseline* decoding scheme. Its uniform rate region $\mathscr{R}^{DS}$ for lower page $X_1$ and upper page $X_2$ is the set of all pairs $(R_1, R_2)$ such that 1) $0 \leqslant R_1 \leqslant I(X_1; Y_1)$ and 2) $0 \leqslant R_2 \leqslant I(X_2; Y_2)$. In $\mathscr{R}^{DS}$, the sum rate $r_s^{DS} = \max\{R_1 + R_2 : (R_1, R_2) \in \mathscr{R}^{DS}\} = I(X_1; Y_1) + I(X_2; Y_2)$.

## IV. PERFORMANCE OF MLC FLASH MEMORY WITH DIFFERENT DECODING SCHEMES AND LABELINGS

In this section, we study the uniform rate region and sum rate of MLC flash memories with different decoding schemes and labelings for different channel models which are characterized by channel transition matrices from voltage levels to quantized outputs. Both a *Program/Erase (P/E) cycling* model and a *data retention* model are considered.

**Definition 3** *For MLC flash memory, the mapping $\sigma_G = (11, 10, 00, 01)$ is called Gray labeling, $\sigma_{NO} = (11, 10, 01, 00)$ is called Natural Order (NO) labeling, and $\sigma_{EO} = (11, 00, 01, 10)$ is called Even Odd (EO) labeling.*

For each labeling, the mapping between inputs $(X_1, X_2) \in \mathcal{T}_{MLC}$ and voltage levels $V \in \mathcal{V}_{MLC}$ is shown in Table I.

### A. Quantization with Three Reads

In this subsection, we fix our quantizer $Q$ with three reads, which are placed between every pair of adjacent voltage levels, as shown in Fig. 1(a). Hence, the output alphabet $\mathcal{Y}_{MLC} =$ $\{s_0, s_1, s_2, s_3\}$. For DS decoding, we assume that the output alphabet for the lower page $X_1$ is $\mathcal{Y}^1_{MLC} = \{s_{0 \cup 1}, s_{2 \cup 3}\}$ of cardinality two, and the output alphabet for the upper page $X_2$ is $\mathcal{Y}^2_{MLC} = \{s_0, s_{1 \cup 2}, s_3\}$ of cardinality three [2]. DS decoding also requires a total of three reads.

We study the performance of MLC flash memories using the *P/E cycling* model, which has different channel characteristics for early and late stages of the memory lifetime.

*1) Early Stage P/E Cycling Model:* The channel transition matrix $p^E_{MLC}(y|v)$, for output $y \in \mathcal{Y}_{MLC}$ and voltage level $v \in \mathcal{V}_{MLC}$, reflects empirical results in [19] and is shown in Table I, where $a_1$, $1 - a_1$, $b_1$, $1 - b_1$, $c_1$, and $1 - c_1$ represent non-zero transition probabilities.

As shown in Table I, note that the *transition probability* of inputs $(X_1, X_2)$ to output $Y$ is determined by 1) the *labeling* which maps inputs to voltage levels, and 2) the *channel transition matrix* from voltage levels to output.

**Lemma 4.** *For channel transition matrix $p^E_{MLC}(y|v)$, using Gray labeling, we have $r^{TIN}_s = r^{SC}_s$ and $\mathcal{R}^{TIN} = \mathcal{R}^{SC}$. Using either NO labeling or EO labeling, we have $r^{TIN}_s < r^{SC}_s$.*

*Proof:* With Gray labeling, in Table I, for the column $Y = s_0$, we have $p_{0,s_0} = 0$, $p_{1,s_0} = 0$, $p_{2,s_0} = 0$, and $p_{3,s_0} = a_1$. Thus, $p_{3,s_0} p_{0,s_0} = p_{2,s_0} p_{1,s_0}$. We can also verify $p_{3,s_i} p_{0,s_i} = p_{2,s_i} p_{1,s_i}$ for $i = 1, 2, 3$. Thus, from Theorem 1, we conclude $r^{TIN}_s = r^{SC}_s$ and $\mathcal{R}^{TIN} = \mathcal{R}^{SC}$. On the other hand, under NO labeling $p_{3,s_2} p_{0,s_2} \neq p_{2,s_2} p_{1,s_2}$, and under EO labeling $p_{3,s_3} p_{0,s_3} \neq p_{2,s_3} p_{1,s_3}$. Thus, from Theorem 1, for these two labelings, $r^{TIN}_s < r^{SC}_s$. ∎

Next, we calculate uniform rate regions and sum rates of the three decoding schemes under different labelings. The results are shown in Table II, where $\lambda_1$, $\lambda_2$, $\lambda_3$, $\lambda_4$, and $\lambda_5$ are

$$\lambda_1 = \frac{f(1-b_1) - f(3-b_1)}{4} + \frac{3}{2},$$

$$\lambda_2 = 1 + \frac{1}{4}\Big(f(1-a_1) + f(1+c_1) + f(1-c_1)\Big) - \frac{1}{4}\Big(f(2-c_1) + f(2-a_1+c_1)\Big),$$

$$\lambda_3 = 1 + \frac{1}{4}\Big(f(1-b_1) + f(c_1) - f(1-b_1+c_1)\Big),$$

$$\lambda_4 = 1 + \frac{1}{4}\Big(f(1-a_1) + f(b_1) + f(1-c_1)\Big) - \frac{1}{4}\Big(f(1-a_1+b_1) + f(2-c_1)\Big),$$

$$\lambda_5 = 1 - \frac{1}{4}\Big(f(1-a_1+b_1) + f(2-c_1) + f(1-b_1+c_1)\Big) + \frac{1}{4}\Big(f(1-a_1) + f(c_1) + f(1-c_1) + f(b_1) + f(1-b_1)\Big).$$

From Table II, we have the following comparisons for different labelings.

**Theorem 5.** *With channel transition matrix $p^E_{MLC}(y|v)$, the rate regions satisfy $\mathcal{R}^{DS}_G \subset \mathcal{R}^{TIN}_G$, $\mathcal{R}^{TIN}_{NO} \subset \mathcal{R}^{TIN}_G$, and $\mathcal{R}^{SC}_G \subset \mathcal{R}^{SC}_{NO}$. For the sum rates, we have $r^{TIN}_{s(G)} > r^{DS}_{s(G)}$, $r^{TIN}_{s(G)} > r^{TIN}_{s(NO)}$, $r^{TIN}_{s(G)} > r^{TIN}_{s(EO)}$, and $r^{SC}_{s(G)} = r^{SC}_{s(NO)} = r^{SC}_{s(EO)}$.*

---

[2] We use the notation $s_{u \cup v}$ to represent an output by merging two outputs $s_u$ and $s_v$ in $\mathcal{Y}_{MLC}$, i.e., $P(Y = s_{u \cup v}|X_1 = x_1, X_2 = x_2) = \sum_{i \in \{u,v\}} P(Y = s_i|X_1 = x_1, X_2 = x_2)$ for any $x_1, x_2 \in \{0, 1\}$. Strictly speaking, for the upper page decoding, current MLC flash memories use an output alphabet $\mathcal{Y}^2_{MLC} = \{s_{1 \cup 2}, s_{0 \cup 3}\}$. The resulting performance cannot exceed that obtained with the output alphabet $\{s_0, s_{1 \cup 2}, s_3\}$ used in this paper.

TABLE II
Uniform rate regions and sum rates of DS, TIN, and SC decodings at early stage of $P/E$ cycling for MLC flash memories

| | | | | |
|---|---|---|---|---|
| Gray | DS | $\mathscr{R}_G^{DS}$ | $0 \leqslant R_1 \leqslant \lambda_1,\ 0 \leqslant R_2 \leqslant \lambda_2$ | $r_{s(G)}^{DS} = \lambda_1 + \lambda_2$ |
| | TIN | $\mathscr{R}_G^{TIN}$ | $0 \leqslant R_1 \leqslant \lambda_3,\ 0 \leqslant R_2 \leqslant \lambda_4$ | $r_{s(G)}^{TIN} = \lambda_3 + \lambda_4$ |
| | SC | $\mathscr{R}_G^{SC}$ | $0 \leqslant R_1 \leqslant \lambda_3,\ 0 \leqslant R_2 \leqslant \lambda_4$ | $r_{s(G)}^{SC} = \lambda_3 + \lambda_4$ |
| NO | TIN | $\mathscr{R}_{NO}^{TIN}$ | $0 \leqslant R_1 \leqslant \lambda_3,\ 0 \leqslant R_2 \leqslant \lambda_5$ | $r_{s(NO)}^{TIN} = \lambda_3 + \lambda_5$ |
| | SC | $\mathscr{R}_{NO}^{SC}$ | $0 \leqslant R_1 \leqslant 1,\ 0 \leqslant R_2 \leqslant \lambda_4,\ R_1 + R_2 \leqslant 1 + \lambda_5$ | $r_{s(NO)}^{SC} = 1 + \lambda_5$ |
| EO | TIN | $\mathscr{R}_{EO}^{TIN}$ | $0 \leqslant R_1 \leqslant \lambda_4,\ 0 \leqslant R_2 \leqslant \lambda_5$ | $r_{s(EO)}^{TIN} = \lambda_4 + \lambda_5$ |
| | SC | $\mathscr{R}_{EO}^{SC}$ | $0 \leqslant R_1 \leqslant 1,\ 0 \leqslant R_2 \leqslant \lambda_3,\ R_1 + R_2 \leqslant 1 + \lambda_5$ | $r_{s(EO)}^{SC} = 1 + \lambda_5$ |

*Proof:* Using Table II, we only need to show $\lambda_3 > \lambda_1$, $\lambda_4 > \lambda_2$, $\lambda_4 > \lambda_5$, $\lambda_3 > \lambda_5$, $1 > \lambda_3$, and $\lambda_3 + \lambda_4 = 1 + \lambda_5$. We prove them as follows.

We have $\lambda_3 - \lambda_1 = \frac{1}{4}\Big(f(c_1) + f(3 - b_1) - f(1 - b_1 + c_1)\Big) - \frac{1}{2}$. Let $h_1(b_1, c_1) = f(c_1) + f(3 - b_1) - f(1 - b_1 + c_1)$. For $0 < b_1 < 1$ and $0 < c_1 < 1$, we have $\frac{\partial h_1(b_1, c_1)}{\partial b_1} = \log_2(1 - b_1 + c_1) - \log_2(3 - b_1) < 0$. Thus, for $0 < b_1 < 1$ and $0 < c_1 < 1$, $h_1(b_1, c_1) > h_1(b_1 = 1, c_1) = 2$, so $\lambda_3 > \lambda_1$. Similarly, $\lambda_4 - \lambda_2 = \frac{1}{4}\Big(f(b_1) + f(2 - a_1 + c_1) - f(1 - a_1 + b_1) - f(1 + c_1)\Big)$. Let $h_2(a_1, b_1, c_1) = f(b_1) + f(2 - a_1 + c_1) - f(1 - a_1 + b_1) - f(1 + c_1)$. For $0 < a_1 < 1$, $0 < b_1 < 1$, and $0 < c_1 < 1$, we have $\frac{\partial h_2(a_1, b_1, c_1)}{\partial a_1} = \log_2(1 - a_1 + b_1) - \log_2(2 - a_1 + c_1) < 0$. Thus, for $0 < a_1 < 1$, $0 < b_1 < 1$, and $0 < c_1 < 1$, $h_2(a_1, b_1, c_1) > h_2(a_1 = 1, b_1, c_1) = 0$, so $\lambda_4 > \lambda_2$.

For $0 < b_1 < 1$ and $0 < c_1 < 1$, we have $f(1 - b_1) + f(c_1) - f(1 - b_1 + c_1) = (1 - b_1)\log_2\frac{1 - b_1}{1 - b_1 + c_1} + c_1\log_2\frac{c_1}{1 - b_1 + c_1} < 0$, so $\lambda_4 - \lambda_5 = -\frac{1}{4}\Big(f(1 - b_1) + f(c_1) - f(1 - b_1 + c_1)\Big) > 0$, and $1 - \lambda_3 = -\frac{1}{4}\Big(f(1 - b_1) + f(c_1) - f(1 - b_1 + c_1)\Big) > 0$.

Finally, $\lambda_3 - \lambda_5 = \frac{1}{4}\Big(f(1 - a_1 + b_1) + f(2 - c_1) - f(1 - a_1) - f(1 - c_1) - f(b_1)\Big)$. Let $h_3(a_1, b_1, c_1) = f(1 - a_1 + b_1) + f(2 - c_1) - f(1 - a_1) - f(1 - c_1) - f(b_1)$. For $0 < a_1 < 1$, $0 < b_1 < 1$, and $0 < c_1 < 1$, we have $\frac{\partial h_3(a_1, b_1, c_1)}{\partial a_1} = \log_2(1 - a_1) - \log_2(1 - a_1 + b_1) < 0$. Thus, for $0 < a_1 < 1$, $0 < b_1 < 1$, and $0 < c_1 < 1$, $h_3(a_1, b_1, c_1) > h_3(a_1 = 1, b_1, c_1) = f(2 - c_1) - f(1 - c_1) > 0$, so $\lambda_3 > \lambda_5$.

The equality $\lambda_3 + \lambda_4 = 1 + \lambda_5$ is obvious. ∎

**Example 1** For the early stage P/E cycling model in Table I, let $a_1 = 0.98$, $b_1 = 0.97$, and $c_1 = 0.99$ [3]. The uniform rate regions under Gray and NO labelings are plotted in Fig. 2(a). It is shown that $\mathscr{R}_{NO}^{TIN} \subset \mathscr{R}_G^{TIN} = \mathscr{R}_G^{SC} \subset \mathscr{R}_{NO}^{SC}$. The SC decoding with NO labeling gives the largest rate region.

---

[3] Since the channel transition matrix varies from different flash chip vendors, the channel parameters are chosen to help visualize the relationship among the rate regions. For other choices of channel parameters, the relative positions of the rate regions and qualitative conclusions stay the same.

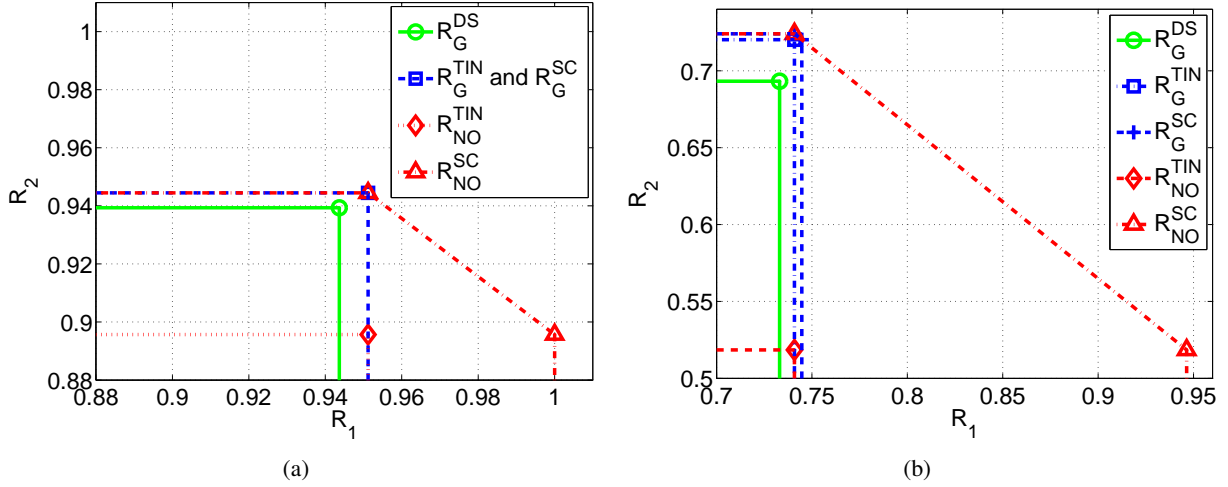

(a)                                                                      (b)

Fig. 2.   (a) Uniform rate regions under Gray and NO labelings with $a_1 = 0.98$, $b_1 = 0.97$, and $c_1 = 0.99$ for the early stage P/E cycling model. (b) Uniform rate regions under Gray and NO labelings with $\hat{a}_1 = 0.82$, $\hat{a}_2 = 0.1$, $\hat{b}_1 = 0.85$, and $\hat{c}_1 = 0.85$ for the late stage P/E cycling model.

TABLE III

CHANNEL TRANSITION MATRIX $p^L_{MLC}(y|v)$ AT LATE STAGE OF $P/E$ CYCLING FOR MLC FLASH MEMORIES

| $V$ | Inputs: $(X_1, X_2)$ | | | Output: $Y$ | | | |
|-----|------|------|------|-------|-------|-------|-------|
| Levels | Gray | NO | EO | $s_0$ | $s_1$ | $s_2$ | $s_3$ |
| $A_0$ | (11) | (11) | (11) | $\hat{a}_1$ | $\hat{a}_2$ | 0 | $1 - \hat{a}_1 - \hat{a}_2$ |
| $A_1$ | (10) | (10) | (00) | 0 | $\hat{b}_1$ | $1 - \hat{b}_1$ | 0 |
| $A_2$ | (00) | (01) | (01) | 0 | 0 | $\hat{c}_1$ | $1 - \hat{c}_1$ |
| $A_3$ | (01) | (00) | (10) | 0 | 0 | 0 | 1 |

TABLE IV

UNIFORM RATE REGIONS AND SUM RATES OF DS, TIN, AND SC DECODINGS AT LATE STAGE OF $P/E$ CYCLING FOR MLC
FLASH MEMORIES

| | | | |
|---|---|---|---|
| Gray | DS | $\mathscr{R}^{DS}_G$ | $0 \leqslant R_1 \leqslant \tau_1, 0 \leqslant R_2 \leqslant \tau_2$ ・ $r^{DS}_{s(G)} = \tau_1 + \tau_2$ |
| | TIN | $\mathscr{R}^{TIN}_G$ | $0 \leqslant R_1 \leqslant \tau_3, 0 \leqslant R_2 \leqslant \tau_4$ ・ $r^{TIN}_{s(G)} = \tau_3 + \tau_4$ |
| | SC | $\mathscr{R}^{SC}_G$ | $0 \leqslant R_1 \leqslant \tau_5, 0 \leqslant R_2 \leqslant \tau_6, R_1 + R_2 \leqslant \tau_4 + \tau_5$ ・ $r^{SC}_{s(G)} = \tau_4 + \tau_5$ |
| NO | TIN | $\mathscr{R}^{TIN}_{NO}$ | $0 \leqslant R_1 \leqslant \tau_3, 0 \leqslant R_2 \leqslant \tau_7$ ・ $r^{TIN}_{s(NO)} = \tau_3 + \tau_7$ |
| | SC | $\mathscr{R}^{SC}_{NO}$ | $0 \leqslant R_1 \leqslant \tau_8, 0 \leqslant R_2 \leqslant \tau_6, R_1 + R_2 \leqslant \tau_7 + \tau_8$ ・ $r^{SC}_{s(NO)} = \tau_7 + \tau_8$ |
| EO | TIN | $\mathscr{R}^{TIN}_{EO}$ | $0 \leqslant R_1 \leqslant \tau_4, 0 \leqslant R_2 \leqslant \tau_7$ ・ $r^{TIN}_{s(EO)} = \tau_4 + \tau_7$ |
| | SC | $\mathscr{R}^{SC}_{EO}$ | $0 \leqslant R_1 \leqslant \tau_8, 0 \leqslant R_2 \leqslant \tau_5, R_1 + R_2 \leqslant \tau_7 + \tau_8$ ・ $r^{SC}_{s(EO)} = \tau_7 + \tau_8$ |

*2) Late Stage P/E Cycling Model:* The channel transition matrix $p^L_{MLC}(y|v)$, for output $y \in \mathcal{Y}_{MLC}$ and voltage level $v \in \mathcal{V}_{MLC}$, reflects measurements in [19] and its structure is shown in Table III where $\hat{a}_1$, $\hat{a}_2$, $1 - \hat{a}_1 - \hat{a}_2$, $\hat{b}_1$, $1 - \hat{b}_1$, $\hat{c}_1$, and $1 - \hat{c}_1$ represent non-zero transition probabilities.

**Lemma 6.** *For channel transition matrix $p^L_{MLC}(y|v)$, we have $r^{TIN}_s < r^{SC}_s$ with Gray labeling, NO labeling or EO labeling.*

*Proof:* For any of Gray labeling, NO labeling or EO labeling, consider the column $Y = s_3$ in Table III. Since three of $p_{0,s_3}$, $p_{1,s_3}$, $p_{2,s_3}$, and $p_{3,s_3}$ are positive, it is impossible to make $p_{3,s_3} p_{0,s_3} = p_{2,s_3} p_{1,s_3} = 0$. Thus, from Theorem 1, we conclude $r^{TIN}_s < r^{SC}_s$. ∎

Next, we calculate uniform rate regions and sum rates of the three decoding schemes under different labelings. The results are shown in Table IV, where $\tau_i$, $i = 1, \ldots, 8$, are

$$\tau_1 = \frac{f(2 - \hat{a}_1 - \hat{a}_2 - \hat{b}_1) - f(4 - \hat{a}_1 - \hat{a}_2 - \hat{b}_1)}{4} + \frac{3}{2},$$

$$\tau_2 = 1 + \frac{1}{4}\Big(f(\hat{a}_2) + f(2 - \hat{a}_1 - \hat{a}_2) + f(1 + \hat{c}_1) + f(1 - \hat{c}_1)\Big) - \frac{1}{4}\Big(f(3 - \hat{a}_1 - \hat{a}_2 - \hat{c}_1) + f(\hat{a}_2 + \hat{c}_1 + 1)\Big),$$

$$\tau_3 = 1 + \frac{1}{4}\Big(f(1 - \hat{b}_1) + f(\hat{c}_1) + f(1 - \hat{a}_1 - \hat{a}_2) + f(2 - \hat{c}_1)\Big) - \frac{1}{4}\Big(f(1 - \hat{b}_1 + \hat{c}_1) + f(3 - \hat{a}_1 - \hat{a}_2 - \hat{c}_1)\Big),$$

$$\tau_4 = 1 + \frac{1}{4}\Big(f(\hat{a}_2) + f(2 - \hat{a}_1 - \hat{a}_2) + f(\hat{b}_1) + f(1 - \hat{c}_1)\Big) - \frac{1}{4}\Big(f(\hat{a}_2 + \hat{b}_1) + f(3 - \hat{a}_1 - \hat{a}_2 - \hat{c}_1)\Big),$$

$$\tau_5 = 1 + \frac{1}{4}\Big(f(1 - \hat{a}_1 - \hat{a}_2) + f(1 - \hat{b}_1) + f(\hat{c}_1)\Big) - \frac{1}{4}\Big(f(2 - \hat{a}_1 - \hat{a}_2) + f(1 - \hat{b}_1 + \hat{c}_1)\Big),$$

$$\tau_6 = 1 + \frac{1}{4}\Big(f(\hat{a}_2) + f(\hat{b}_1) + f(1 - \hat{c}_1)\Big) - \frac{1}{4}\Big(f(\hat{a}_2 + \hat{b}_1) + f(2 - \hat{c}_1)\Big),$$

$$\tau_7 = 1 - \frac{1}{4}\Big(f(\hat{a}_2 + \hat{b}_1) + f(1 - \hat{b}_1 + \hat{c}_1) + f(3 - \hat{a}_1 - \hat{a}_2 - \hat{c}_1)\Big)$$
$$\qquad + \frac{1}{4}\Big(f(\hat{a}_2) + f(\hat{c}_1) + f(2 - \hat{a}_1 - \hat{a}_2 - \hat{c}_1) + f(\hat{b}_1) + f(1 - \hat{b}_1)\Big),$$

$$\tau_8 = 1 + \frac{1}{4}\Big(f(1 - \hat{c}_1) + f(1 - \hat{a}_1 - \hat{a}_2) - f(2 - \hat{a}_1 - \hat{a}_2 - \hat{c}_1)\Big).$$

From Table IV, we have the following comparisons for different labelings.

**Theorem 7.** *With channel transition matrix $p_{MLC}^L(y|v)$, the rate regions satisfy $\mathscr{R}_G^{DS} \subset \mathscr{R}_G^{TIN}$, $\mathscr{R}_{NO}^{TIN} \subset \mathscr{R}_G^{TIN}$, and $\mathscr{R}_G^{SC} \subset \mathscr{R}_{NO}^{SC}$. For the sum rates, we have $r_{s(G)}^{TIN} > r_{s(G)}^{DS}$, $r_{s(G)}^{TIN} > r_{s(NO)}^{TIN}$, $r_{s(G)}^{TIN} > r_{s(EO)}^{TIN}$, and $r_{s(G)}^{SC} = r_{s(NO)}^{SC} = r_{s(EO)}^{SC}$.*

*Proof:* Using Table IV, we only need to show $\tau_3 \geqslant \tau_1$, $\tau_4 > \tau_2$, $\tau_4 > \tau_7$, $\tau_8 > \tau_5$, $\tau_3 > \tau_7$, and $\tau_4 + \tau_5 = \tau_7 + \tau_8$. Here, we give proofs of $\tau_3 \geqslant \tau_1$ and $\tau_4 > \tau_2$. Others can be proved in a similar way.

$$4(\tau_3 - \tau_1) = f(1 - \hat{b}_1) + f(\hat{c}_1) + f(1 - \hat{a}_1 - \hat{a}_2) + f(2 - \hat{c}_1) + f(4 - \hat{a}_1 - \hat{a}_2 - \hat{b}_1)$$
$$\qquad - f(1 - \hat{b}_1 + \hat{c}_1) - f(3 - \hat{a}_1 - \hat{a}_2 - \hat{c}_1) - f(2 - \hat{a}_1 - \hat{a}_2 - \hat{b}_1) - f(2)$$
$$\qquad = (1 - \hat{b}_1 + \hat{c}_1)\log_2(1 + \frac{3 - \hat{a}_1 - \hat{a}_2 - \hat{c}_1}{1 - \hat{b}_1 + \hat{c}_1}) - (1 - \hat{b}_1)\log_2(1 + \frac{1 - \hat{a}_1 - \hat{a}_2}{1 - \hat{b}_1}) - \hat{c}_1\log_2(1 + \frac{2 - \hat{c}_1}{\hat{c}_1})$$
$$\qquad + (3 - \hat{a}_1 - \hat{a}_2 - \hat{c}_1)\log_2(1 + \frac{1 - \hat{b}_1 + \hat{c}_1}{3 - \hat{a}_1 - \hat{a}_2 - \hat{c}_1}) - (1 - \hat{a}_1 - \hat{a}_2)\log_2(1 + \frac{1 - \hat{b}_1}{1 - \hat{a}_1 - \hat{a}_2})$$
$$\qquad - (2 - \hat{c}_1)\log_2(1 + \frac{\hat{c}_1}{2 - \hat{c}_1}).$$

The function $t\log_2(1 + 1/t)$ is concave. Let $t_1 = \frac{1 - \hat{b}_1}{1 - \hat{a}_1 - \hat{a}_2}$, $t_2 = \frac{\hat{c}_1}{2 - \hat{c}_1}$, $r_1 = \frac{1 - \hat{a}_1 - \hat{a}_2}{3 - \hat{a}_1 - \hat{a}_2 - \hat{c}_1}$, and $r_2 = \frac{2 - \hat{c}_1}{3 - \hat{a}_1 - \hat{a}_2 - \hat{c}_1}$. Then we have $(r_1 t_1 + r_2 t_2)\log_2\big(1 + 1/(r_1 t_1 + r_2 t_2)\big) \geqslant r_1 t_1 \log_2(1 + 1/t_1) + r_2 t_2 \log_2(1 + 1/t_2)$; that is, $(1 - \hat{b}_1 + \hat{c}_1)\log_2(1 + \frac{3 - \hat{a}_1 - \hat{a}_2 - \hat{c}_1}{1 - \hat{b}_1 + \hat{c}_1}) \geqslant (1 - \hat{b}_1)\log_2(1 + \frac{1 - \hat{a}_1 - \hat{a}_2}{1 - \hat{b}_1}) + \hat{c}_1\log_2(1 + \frac{2 - \hat{c}_1}{\hat{c}_1})$. Similarly, we have $(3 - \hat{a}_1 - \hat{a}_2 - \hat{c}_1)\log_2(1 + \frac{1 - \hat{b}_1 + \hat{c}_1}{3 - \hat{a}_1 - \hat{a}_2 - \hat{c}_1}) \geqslant (1 - \hat{a}_1 - \hat{a}_2)\log_2(1 + \frac{1 - \hat{b}_1}{1 - \hat{a}_1 - \hat{a}_2}) + (2 - \hat{c}_1)\log_2(1 + \frac{\hat{c}_1}{2 - \hat{c}_1})$. Therefore, $\tau_3 - \tau_1 \geqslant 0$.

Next, $\tau_4 - \tau_2 = \frac{1}{4}\Big(f(\hat{b}_1) + f(\hat{a}_2 + \hat{c}_1 + 1) - f(\hat{a}_2 + \hat{b}_1) - f(1 + \hat{c}_1)\Big)$. Let $\hat{h}_1(\hat{a}_2, \hat{b}_1, \hat{c}_1) = f(\hat{b}_1) + f(\hat{a}_2 + \hat{c}_1 + 1) - f(\hat{a}_2 + \hat{b}_1) - f(1 + \hat{c}_1)$. For $0 < \hat{a}_2 < 1$, $0 < \hat{b}_1 < 1$, and $0 <$

$\hat{c}_1 < 1$, we have $\frac{\partial \hat{h}_1(\hat{a}_2, \hat{b}_1, \hat{c}_1)}{\partial \hat{a}_2} = \log_2(\hat{a}_2 + \hat{c}_1 + 1) - \log_2(\hat{a}_2 + \hat{b}_1) > 0$. Thus, for $0 < \hat{a}_2 < 1$, $0 < \hat{b}_1 < 1$, and $0 < \hat{c}_1 < 1$, $\hat{h}_1(\hat{a}_2, \hat{b}_1, \hat{c}_1) > \hat{h}_1(\hat{a}_2 = 0, \hat{b}_1, \hat{c}_1) = 0$, so $\tau_4 > \tau_2$. ∎

**Example 2** For the late stage P/E cycling model in Table III, let $\hat{a}_1 = 0.82$, $\hat{a}_2 = 0.1$, $\hat{b}_1 = 0.85$, and $\hat{c}_1 = 0.85$. The uniform rate regions under Gray and NO labelings are plotted in Fig. 2(b). It is shown that $\mathscr{R}_{NO}^{TIN} \subset \mathscr{R}_{G}^{TIN} \subset \mathscr{R}_{G}^{SC} \subset \mathscr{R}_{NO}^{SC}$. Note that unlike the early stage P/E cycling model in Example 1, here the region $\mathscr{R}_{G}^{TIN}$ is strictly included in $\mathscr{R}_{G}^{SC}$.

**Remark 2** For the P/E cycling model (both early and late stages), from Theorems 5 and 7, for TIN decoding, among the 3 labelings, Gray labeling gives the largest sum rate which is larger than that of DS decoding. Moreover, compared to NO labeling, Gray labeling generates a larger uniform rate region for TIN decoding, but a smaller one for SC decoding.

For the early stage P/E cycling model, from Lemma 4, Theorem 5, and Table II, we have the following two main observations.

1) The sum rate of TIN decoding under Gray labeling is the same as that of SC decoding under any of Gray, NO, and EO labelings. This provides an ECC solution for MLC flash based on good codes for the point-to-point channel. With Gray labeling, we only need to use two point-to-point capacity-achieving codes, e.g., polar codes [1], for the lower and upper pages, to achieve the rates $I(X_1; Y)$ and $I(X_2; Y)$, respectively. For the decoding, the two pages are just decoded with polar code decoders independently.

2) For NO labeling or EO labeling, with SC decoding, rate $(R_1 = 1, R_2 = \lambda_5)$ can be achieved, which means that the lower page $X_1$ does not need coding. This also gives us a very simple ECC solution for MLC flash. For encoding, we only need to apply a point-to-point capacity-achieving code, e.g., polar code [1], to the upper page $X_2$ to achieve rate $I(X_2; Y)$, and no coding is needed for the lower page $X_1$. For decoding, we first decode the upper page. Then, based on the decoded data from the upper page, binary labeling, channel transition matrix, and output $Y$, the data of the lower page can be determined, e.g., with NO labeling in Table I, if the correctly decoded bit of the upper page is $X_2 = 1$ and the output $Y = s_2$, then the lower page bit is determined as $X_1 = 0$.

For the late stage P/E cycling model, from Lemma 6, Theorem 7, and Table IV, the sum rate of TIN decoding under Gray labeling is strictly less than that of SC decoding. The gap $\Delta$ between the two is

$$\Delta = r_{s(G)}^{SC} - r_{s(G)}^{TIN} = \tau_5 - \tau_3 = \frac{1}{4}\Big(f(3 - \hat{a}_1 - \hat{a}_2 - \hat{c}_1) - f(2 - \hat{a}_1 - \hat{a}_2) - f(2 - \hat{c}_1)\Big).$$

To bound $\Delta$, let $\hat{a} = \hat{a}_1 + \hat{a}_2$ and $\hat{h}(\hat{a}, \hat{c}_1) = f(3 - \hat{a} - \hat{c}_1) - f(2 - \hat{a}) - f(2 - \hat{c}_1)$. For $0 < \hat{a} < 1$ and $0 < \hat{c}_1 < 1$, we have $\frac{\partial \hat{h}(\hat{a}, \hat{c}_1)}{\partial \hat{a}} = \log_2(2 - \hat{a}) - \log_2(3 - \hat{a} - \hat{c}_1) < 0$, and $\frac{\partial \hat{h}(\hat{a}, \hat{c}_1)}{\partial \hat{c}_1} = \log_2(2 - \hat{c}_1) - \log_2(3 - \hat{a} - \hat{c}_1) < 0$. Therefore, $\frac{\hat{h}(\hat{a}=1, \hat{c}_1=1)}{4} < \Delta < \frac{\hat{h}(\hat{a}=0, \hat{c}_1=0)}{4}$; that is, $0 <$ $\Delta < \frac{3\log_2 3 - 4}{4} = 0.1887$. If we impose constraints $\eta_{\hat{a}} \leqslant \hat{a}_1 + \hat{a}_2 < 1$ and $\eta_{\hat{c}_1} \leqslant \hat{c}_1 < 1$, we have $0 < \Delta \leqslant \frac{1}{4}\Big(f(3-\eta_{\hat{a}}-\eta_{\hat{c}_1}) - f(2-\eta_{\hat{a}}) - f(2-\eta_{\hat{c}_1})\Big)$. For example, for $\eta_{\hat{a}} = 0.95$ and $\eta_{\hat{c}_1} = 0.85$, we get $0 < \Delta \leqslant 0.00246$. In general, the gap $\Delta$ is very small. $\qquad\square$

TABLE V
Channel transition matrix $p_{MLC}^{DR}(y|v)$ of data retention model for MLC flash memories.

| $V$ | Inputs: $(X_1, X_2)$ | | | Output: $Y$ | | | |
|-----|------|------|------|------|------|------|------|
| Levels | Gray | NO | EO | $s_0$ | $s_1$ | $s_2$ | $s_3$ |
| $A_0$ | (11) | (11) | (11) | 1 | 0 | 0 | 0 |
| $A_1$ | (10) | (10) | (00) | $1-\tilde{a}_1$ | $\tilde{a}_1$ | 0 | 0 |
| $A_2$ | (00) | (01) | (01) | 0 | $1-\tilde{b}_1$ | $\tilde{b}_1$ | 0 |
| $A_3$ | (01) | (00) | (10) | 0 | 0 | $1-\tilde{c}_1$ | $\tilde{c}_1$ |

For the *data retention* model, the channel transition matrix $p_{MLC}^{DR}(y|v)$, for output $y \in \mathcal{Y}_{MLC}$ and voltage level $v \in \mathcal{V}_{MLC}$, reflects measured data and its structure is shown in Table V where $\tilde{a}_1$, $1-\tilde{a}_1$, $\tilde{b}_1$, $1-\tilde{b}_1$, $\tilde{c}_1$, and $1-\tilde{c}_1$ represent non-zero transition probabilities. Compared to the early stage P/E cycling model where errors are caused by cell voltage upward drift, in this data retention model, errors are due to the cell voltage reduction. Analysis and results of this model are very similar to the early stage P/E cycling model. We only give one result here.

**Lemma 8.** *For channel transition matrix $p_{MLC}^{DR}(y|v)$, using Gray labeling, we have $r_s^{TIN} = r_s^{SC}$ and $\mathscr{R}^{TIN} = \mathscr{R}^{SC}$. Using either NO labeling or EO labeling, we have $r_s^{TIN} < r_s^{SC}$.*

*Proof:* The proof is omitted, since it is similar to that of Lemma 4. $\qquad\blacksquare$

Next, we study the structure and property of all labelings. There exist a total of $4! = 24$ labelings. In order to categorize and analyze these 24 labelings, we take advantage of the algebraic structure of groups, and consider a *labeling* $\sigma$ as a *permutation* $\pi$ in the *symmetric group* $\mathcal{S}_4$. This is the group whose elements are all the permutation operations that can be performed on 4 distinct elements in $\mathcal{T}_{MLC}$, and whose group operation, denoted as $*$, is the composition of such permutation operations, which are defined as bijective functions from the set $\mathcal{T}_{MLC}$ to itself. A labeling $\sigma = (w_0, w_1, w_2, w_3)$ corresponds to the permutation $\pi = (w_0, w_1, w_2, w_3)$ in $\mathcal{S}_4$, where the permutation vector $\pi = (w_0, w_1, w_2, w_3)$ (the $w_i$, $i = 0, 1, 2, 3$, represent the full set of possible 2-tuples) is defined to represent $\pi(11) = w_0$, $\pi(10) = w_1$, $\pi(01) = w_2$, and $\pi(00) = w_3$, e.g., $\pi = (11, 10, 01, 00)$ is the identity permutation in $\mathcal{S}_4$. The group operation $*$ of two permutations $\pi_1$ and $\pi_2$ is defined as their composition and results in another permutation $\pi_3 = \pi_1 * \pi_2$. In other words, $\pi_1 * \pi_2$ is the function that maps any element $w \in \mathcal{T}_{MLC}$ to $\pi_1\big(\pi_2(w)\big)$. Note that the rightmost permutation is applied first. For example, $(10, 11, 00, 01) * (11, 10, 00, 01) = (10, 11, 01, 00)$.

**Lemma 9.** *In the symmetric group $\mathcal{S}_4$, $G_0 = \{(11, 10, 01, 00), (10, 11, 00, 01), (01, 00, 11, 10), (00, 01, 10, 11)\}$ forms a normal subgroup (the Klein four-group).*

*Proof:* The element $(11, 10, 01, 00)$ in $G_0$ is the identity element in $\mathcal{S}_4$. We can verify 4 elements in $G_0$ have the following properties: 1) the composition of the identity element and any element is that element itself; 2) the composition of any non-identity element with itself is the identity element; 3) the composition of two distinct non-identity elements is the third non-identity element. Thus, $G_0$ is the Klein four-group. ∎

With the subgroup $G_0$ in Lemma 9, we partition $\mathcal{S}_4$ into $G_0$ and its 5 cosets, each of size 4: $\mathcal{S}_4 = G_0 \cup G_1 \cup G_2 \cup \bar{G}_0 \cup \bar{G}_1 \cup \bar{G}_2$, where $G_1 = G_0 * (11, 10, 00, 01)$; $G_2 = G_0 * (11, 00, 01, 10)$; $\bar{G}_0 = G_0 * (11, 01, 10, 00)$; $\bar{G}_1 = G_0 * (11, 01, 00, 10)$; $\bar{G}_2 = G_0 * (11, 00, 10, 01)$.

In the following, we will treat each vector in every coset as a labeling. For example, $G_0$ includes the NO labeling $\sigma_{NO} = (11, 10, 01, 00)$, $G_1$ includes $\sigma_G = (11, 10, 00, 01)$, and $G_2$ includes $\sigma_{EO} = (11, 00, 01, 10)$. The following two lemmas give properties of the uniform rate regions for different labelings. We assume an *arbitrary* channel transition matrix $p_{MLC}(y|v)$ where output $y \in \mathcal{Y}_{MLC}$ and voltage level $v \in \mathcal{V}_{MLC}$ is given. The following lemma leverages the symmetries within the Klein four-group and its cosets to deduce the relationship of the rate regions of different labelings.

**Lemma 10.** *With an arbitrary channel transition matrix $p_{MLC}(y|v)$, for TIN decoding, the 4 labelings in each of $G_0$, $G_1$, $G_2$, $\bar{G}_0$, $\bar{G}_1$, and $\bar{G}_2$ give the same uniform rate region $\mathscr{R}^{TIN}$ and sum rate $r_s^{TIN}$. For SC decoding, the 4 labelings in each of $G_0$, $G_1$, $G_2$, $\bar{G}_0$, $\bar{G}_1$, and $\bar{G}_2$ give the same uniform rate region $\mathscr{R}^{SC}$, and all 24 labelings in $\mathcal{S}_4$ give the same sum rate $r_s^{SC}$.*

*Proof:* This is based on the fact that 4 labelings in each of $G_0$, $G_1$, $G_2$, $\bar{G}_0$, $\bar{G}_1$, and $\bar{G}_2$ are interchangeable by one of the following three operations: 1) on position $X_1$, change 0 to 1 and 1 to 0; 2) on position $X_2$, change 0 to 1 and 1 to 0; 3) on both positions $X_1$ and $X_2$, change 0 to 1 and 1 to 0. For example, in $G_0$, $(11, 10, 01, 00)$ is transformed to $(01, 00, 11, 10)$ by changing 0 to 1 and 1 to 0 on position $X_1$, is transformed to $(10, 11, 00, 01)$ by changing 0 to 1 and 1 to 0 on position $X_2$, and is transformed to $(00, 01, 10, 11)$ by changing 0 to 1 and 1 to 0 on both positions $X_1$ and $X_2$. Since the input distributions for $X_1$ and $X_2$ are uniform, the values of $I(X_1; Y)$, $I(X_2; Y)$, $I(X_1; Y|X_2)$, and $I(X_2; Y|X_1)$ under a labeling $\sigma_1$ are the same as those under a labeling $\sigma_2$ which is obtained by one of the above three operations on the labeling $\sigma_1$. Thus, for a fixed decoding scheme (TIN or SC), the uniform region and sum rate under the labeling $\sigma_1$ are the same as those under the labeling $\sigma_2$. Therefore, for a fixed decoding scheme (TIN or SC), the 4 labelings in each coset give the same uniform rate region and sum rate. For the sum rate of SC decoding, for all 24 labelings, $I(X_1, X_2; Y)$ is the same due to the uniform input distributions for $X_1$ and $X_2$. ∎

**Lemma 11.** *With an arbitrary channel transition matrix $p_{MLC}(y|v)$, for TIN decoding, if labelings in $G_i$, $i = 0, 1, 2$, give a uniform rate region: $0 \leqslant R_1 \leqslant \varphi_1$ and $0 \leqslant R_2 \leqslant \varphi_2$, then labelings in $\bar{G}_i$ give a uniform rate region: $0 \leqslant R_1 \leqslant \varphi_2$ and $0 \leqslant R_2 \leqslant \varphi_1$. For SC decoding, if labelings in $G_i$ give a uniform rate region: $0 \leqslant R_1 \leqslant \psi_1$, $0 \leqslant R_2 \leqslant \psi_2$, and $R_1 + R_2 \leqslant \psi_3$, then labelings in $\bar{G}_i$ give a uniform rate region: $0 \leqslant R_1 \leqslant \psi_2$, $0 \leqslant R_2 \leqslant \psi_1$, and $R_1 + R_2 \leqslant \psi_3$.*

*Proof:* This is based on the fact that 4 labelings in $G_i$, $i=0, 1, 2$, are transformed (one-to-one) to 4 labelings in $\bar{G}_i$ by swapping the values on $X_1$ and $X_2$. For example, $(11, 10, 01, 00)$ in $G_0$ is transformed to $(11, 01, 10, 00)$ in $\bar{G}_0$. With the uniform input distributions for $X_1$ and $X_2$, $X_1$ (or $X_2$) with labeling $(11, 10, 01, 00)$ is equivalent to $X_2$ (or $X_1$) with labeling $(11, 01, 10, 00)$. Thus, for a fixed decoding scheme (TIN or SC), the uniform rate region under labeling $(11, 10, 01, 00)$ will become the one under labeling $(11, 01, 10, 00)$, by swapping the constraints on $R_1$ and $R_2$. Since for a fixed decoding scheme (TIN or SC) the 4 labelings in each coset give the same uniform rate region from Lemma 10, the uniform rate region under labelings in $G_0$ will become the one under labelings in $\bar{G}_0$, by swapping the constraints on $R_1$ and $R_2$. The same conclusion holds for labelings in $G_i$ and $\bar{G}_i$, $i = 1, 2$. ∎

**Remark 3** From Theorems 5 and 7, and Lemmas 10 and 11, under the P/E cycling model (both early and late stages), for TIN decoding, the 8 labelings in $G_1$ (including Gray labeling) and $\bar{G}_1$ among all 24 labelings produce the largest sum rate. For SC decoding, all the 24 labelings give the same sum rate. □

Finally, we discuss the uniform rate region if we are allowed to use multiple labelings together for each codeword instead of one labeling in MLC flash. In current flash memory, only Gray labeling is used. More specifically, we study the uniform rate region achieved by *time sharing* of using labelings in $\mathcal{S}_4$.

Define $\mathscr{R}_{\mathcal{S}_4}^{TIN} = Conv\left( \bigcup_{\sigma \in \mathcal{S}_4} \mathscr{R}_{\sigma}^{TIN} \right)$, the convex hull of uniform rate regions of all 24 labelings for TIN decoding. Define $\mathscr{R}_{\mathcal{S}_4}^{SC} = Conv\left( \bigcup_{\sigma \in \mathcal{S}_4} \mathscr{R}_{\sigma}^{SC} \right)$, the convex hull of uniform rate regions of all 24 labelings for SC decoding. Through time sharing of different labelings, we have the following lemma.

**Lemma 12.** *For TIN decoding, any point $(R_1, R_2) \in \mathscr{R}_{\mathcal{S}_4}^{TIN}$ can be achieved. For SC decoding, any point $(R_1, R_2) \in \mathscr{R}_{\mathcal{S}_4}^{SC}$ can be achieved.*

*Proof:* We first show that any point $(R_1, R_2) \in \mathscr{R}_{\mathcal{S}_4}^{TIN}$ can be achieved. From Carathéodory's Theorem [6], any point $(R_1, R_2)$ in $\mathscr{R}_{\mathcal{S}_4}^{TIN}$ can be represented as a convex combination of 3 points in the $\bigcup_{\sigma \in \mathcal{S}_4} \mathscr{R}^{TIN}$. Without loss of generality, we assume $(R_1, R_2) = \alpha_1(R_1^1, R_2^1) + \alpha_2(R_1^2, R_2^2) + \alpha_3(R_1^3, R_2^3)$, where $\alpha_1, \alpha_2, \alpha_3 \geqslant 0$ and $\sum_{i=1}^{3} \alpha_i = 1$. Points $(R_1^1, R_2^1)$, $(R_1^2, R_2^2)$, and $(R_1^3, R_2^3)$ in $\bigcup_{\sigma \in \mathcal{S}_4} \mathscr{R}_{\sigma}^{TIN}$ are achievable under some labelings. Consider three sequences of codes, achieving $(R_1^1, R_2^1)$, $(R_1^2, R_2^2)$, and $(R_1^3, R_2^3)$, respectively. For each block length $n$, consider the $(2^{\alpha_1 n R_1^1}, 2^{\alpha_1 n R_2^1}, \alpha_1 n)$, $(2^{\alpha_2 n R_1^2}, 2^{\alpha_2 n R_2^2}, \alpha_2 n)$, and $(2^{\alpha_3 n R_1^3}, 2^{\alpha_3 n R_2^3}, \alpha_3 n)$ codes from the given three sequences of codes, respectively. By a standard time sharing argument [6], a fourth $(2^{n R_1}, 2^{n R_2}, n)$ code can be constructed from the above three codes (we omit details here due to space limitation). Thus, any point $(R_1, R_2) \in \mathscr{R}_{\mathcal{S}_4}^{TIN}$ can be achieved. That any point $(R_1, R_2) \in \mathscr{R}_{\mathcal{S}_4}^{SC}$ is achievable can be proved in a similar way. ∎

Moreover, for the early stage P/E cycling model in Table I, the rate region $\mathscr{R}_{\mathcal{S}_4}^{SC}$ can be determined explicitly.

**Theorem 13.** *For the early stage P/E cycling model, $\mathscr{R}_{\mathcal{S}_4}^{SC}$ is the set of all pairs $(R_1, R_2)$ such that $0 \leqslant R_1 \leqslant 1$, $0 \leqslant R_2 \leqslant 1$, and $R_1 + R_2 \leqslant I(X_1, X_2; Y) = 1 + \lambda_5$.*

*Proof:* Using Table II, Lemma 10, and Lemma 11, the convex hull of the uniform rate regions of all 24 labelings can be calculated. We also find the convex hull of the uniform rate regions of two labelings $(11, 10, 01, 00)$ and $(11, 01, 10, 00)$ are enough to achieve $\mathscr{R}_{\mathcal{S}_4}^{SC}$. ∎

**Example 3** For the early stage of P/E cycling, let $a_1 = 0.98$, $b_1 = 0.97$, and $c_1 = 0.99$. The uniform rate regions $\mathscr{R}_{\mathcal{S}_4}^{DS}$, $\mathscr{R}_{\mathcal{S}_4}^{SC}$, and $\mathscr{R}_G^{DS}$ are plotted in Fig. 3. It is shown $\mathscr{R}_G^{DS} \subset \mathscr{R}_{\mathcal{S}_4}^{TIN} \subset \mathscr{R}_{\mathcal{S}_4}^{SC}$, and the line connecting the two corner points in $\mathscr{R}_{\mathcal{S}_4}^{TIN}$ is on the line connecting the two corner points in $\mathscr{R}_{\mathcal{S}_4}^{SC}$. Moreover, either $R_1$ or $R_2$ can achieve rate 1 with SC decoding. For the late stage P/E cycling, let $\hat{a}_1 = 0.82$, $\hat{a}_2 = 0.1$, $\hat{b}_1 = 0.85$, and $\hat{c}_1 = 0.85$. The uniform rate regions $\mathscr{R}_{\mathcal{S}_4}^{TIN}$, $\mathscr{R}_{\mathcal{S}_4}^{SC}$, and $\mathscr{R}_G^{DS}$ are plotted in Fig. 4. For this case, there is a gap between the line connecting the two corner points in $\mathscr{R}_{\mathcal{S}_4}^{TIN}$ and the one connecting the two corner points in $\mathscr{R}_{\mathcal{S}_4}^{SC}$. It means that SC decoding will give larger sum rate than that of TIN decoding.

### B. Quantization with Multiple Reads

In the above subsection, we use a quantizer with three reads. To improve decoding performance, we can progressively apply multiple reads to obtain more soft information.

For a channel $\mathcal{W}_{MLC}$, assume we already have an output set $\mathcal{Y}_{MLC}^q = \{s_0, s_1, \ldots, s_{q-1}\}$ obtained by a set of reads. The corresponding uniform rate regions for TIN and SC decodings are $\mathscr{R}^{TIN}$ and $\mathscr{R}^{SC}$, and the sum rates for TIN and SC decodings are $r_s^{TIN}$ and $r_s^{SC}$.

Now, we apply one more read (quantization) to split one of the outputs $s_0, s_1, \ldots, s_{q-1}$. Without loss of generality, we split $s_0$ into $s_0^1$ and $s_0^2$ to obtain a new output set $\hat{\mathcal{Y}}_{MLC}^{q+1} = \{s_0^1, s_0^2, s_1, s_2, \ldots, s_{q-1}\}$. The resulting uniform rate regions for TIN and SC decodings become $\hat{\mathscr{R}}^{TIN}$ and $\hat{\mathscr{R}}^{SC}$, and the corresponding sum rates become $\hat{r}_s^{TIN}$ and $\hat{r}_s^{SC}$. The following lemma shows that such one-step progressive quantization will not decrease (in general, strictly increase) the performance.

**Lemma 14.** *For uniform rate regions, under TIN and SC decodings, $\mathscr{R}^{TIN} \subseteq \hat{\mathscr{R}}^{TIN}$ and $\mathscr{R}^{SC} \subseteq \hat{\mathscr{R}}^{SC}$. For sum rates, under TIN and SC decodings, $r_s^{TIN} \leqslant \hat{r}_s^{TIN}$ and $r_s^{SC} \leqslant \hat{r}_s^{SC}$.*

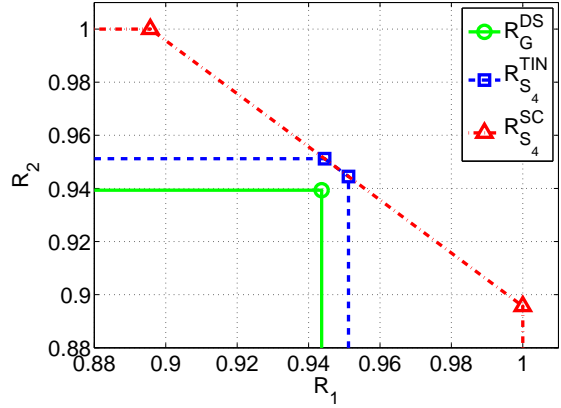

Fig. 3. Uniform rate regions $\mathscr{R}_{S_4}^{TIN}$, $\mathscr{R}_{S_4}^{SC}$, and $\mathscr{R}_{G}^{DS}$ (baseline) with $a_1 = 0.98$, $b_1 = 0.97$, and $c_1 = 0.99$ for the early stage P/E cycling model.

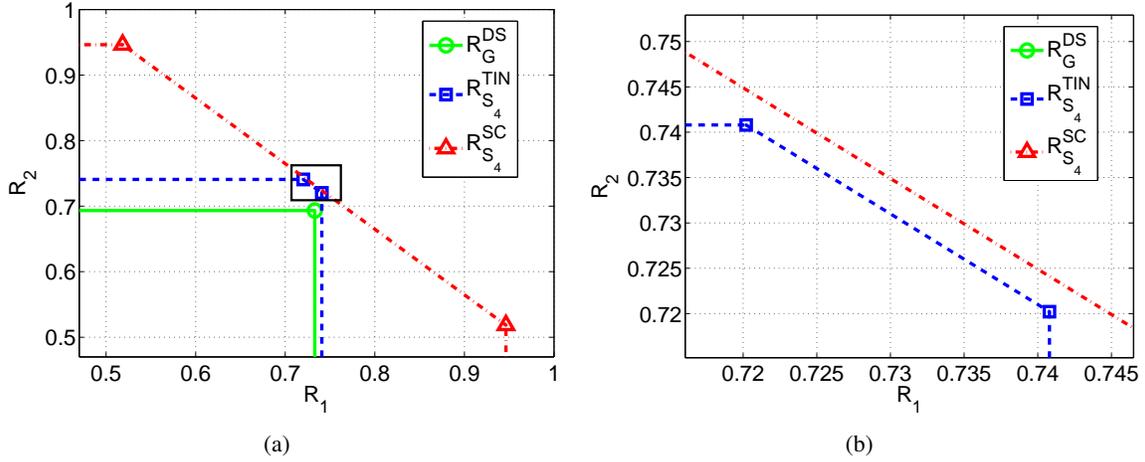

(a)            (b)

Fig. 4. (a) Uniform rate regions $\mathscr{R}_{S_4}^{TIN}$, $\mathscr{R}_{S_4}^{SC}$, and $\mathscr{R}_{G}^{DS}$ (baseline) with $\hat{a}_1 = 0.82$, $\hat{a}_2 = 0.1$, $\hat{b}_1 = 0.85$, and $\hat{c}_1 = 0.85$ for the late stage P/E cycling model, where the two curves (blue and red) in the black rectangle are zoomed in and shown in (b).

*Proof:* We first prove the sum rate $\hat{r}_s^{SC} \geqslant r_s^{SC}$, i.e., $I(X_1, X_2; \hat{Y}) \geqslant I(X_1, X_2; Y)$, and give the condition when the equality holds. The $I(X_1, X_2; Y)$ can be expressed as follows:

$$I(X_1, X_2; Y) = H(Y) - H(Y|X_1, X_2)$$

$$= H\left(\frac{\sum_{i=0}^{3} p_{i,s_0}}{4}, \frac{\sum_{i=0}^{3} p_{i,s_1}}{4}, \ldots, \frac{\sum_{i=0}^{3} p_{i,s_{q-1}}}{4}\right) - \frac{1}{4}\sum_{i=0}^{3} H(p_{i,s_0}, p_{i,s_1}, \ldots, p_{i,s_{q-1}}).$$

Similarly, the $I(X_1, X_2; \hat{Y})$ is

$$I(X_1, X_2; \hat{Y}) = H(\hat{Y}) - H(\hat{Y}|X_1, X_2)$$

$$= H\left(\frac{\sum_{i=0}^{3} p_{i,s_0^1}}{4}, \frac{\sum_{i=0}^{3} p_{i,s_0^2}}{4}, \frac{\sum_{i=0}^{3} p_{i,s_1}}{4}, \ldots, \frac{\sum_{i=0}^{3} p_{i,s_{q-1}}}{4}\right) - \frac{1}{4}\sum_{i=0}^{3} H(p_{i,s_0^1}, p_{i,s_0^2}, p_{i,s_1}, \ldots, p_{i,s_{q-1}}).$$

$$\stackrel{(a)}{=} H\left(\frac{\sum_{i=0}^{3} p_{i,s_0}}{4}, \frac{\sum_{i=0}^{3} p_{i,s_1}}{4}, \ldots, \frac{\sum_{i=0}^{3} p_{i,s_{q-1}}}{4}\right) + \frac{\sum_{i=0}^{3} p_{i,s_0}}{4} H\left(\frac{\sum_{i=0}^{3} p_{i,s_0^1}}{\sum_{i=0}^{3} p_{i,s_0}}, \frac{\sum_{i=0}^{3} p_{i,s_0^2}}{\sum_{i=0}^{3} p_{i,s_0}}\right)$$

$$- \frac{1}{4}\sum_{i=0}^{3} H(p_{i,s_0}, p_{i,s_1}, \ldots, p_{i,s_{q-1}}) - \frac{1}{4}\sum_{i=0}^{3} p_{i,s_0} H\left(\frac{p_{i,s_0^1}}{p_{i,s_0}}, \frac{p_{i,s_0^2}}{p_{i,s_0}}\right),$$

where step (a) is from the grouping property of entropy [15] and $p_{i,s_0} = p_{i,s_0^1} + p_{i,s_0^2}$ for $i = 0, 1, 2, 3$.

The difference between $I(X_1, X_2; \hat{Y})$ and $I(X_1, X_2; Y)$ is

$$I(X_1, X_2; \hat{Y}) - I(X_1, X_2; Y)$$

$$= \frac{\sum_{i=0}^{3} p_{i,s_0}}{4} H\left(\frac{\sum_{i=0}^{3} p_{i,s_0^1}}{\sum_{i=0}^{3} p_{i,s_0}}, \frac{\sum_{i=0}^{3} p_{i,s_0^2}}{\sum_{i=0}^{3} p_{i,s_0}}\right) - \frac{1}{4}\sum_{i=0}^{3} p_{i,s_0} H\left(\frac{p_{i,s_0^1}}{p_{i,s_0}}, \frac{p_{i,s_0^2}}{p_{i,s_0}}\right)$$

$$= \frac{\sum_{i=0}^{3} p_{i,s_0}}{4}\left(-\frac{1}{\sum_{i=0}^{3} p_{i,s_0}}\left(f(\sum_{i=0}^{3} p_{i,s_0^1}) + f(\sum_{i=0}^{3} p_{i,s_0^2})\right) + \log_2(\sum_{i=0}^{3} p_{i,s_0})\right)$$

$$\quad - \frac{1}{4}\sum_{i=0}^{3} p_{i,s_0}\left(-\frac{1}{p_{i,s_0}}\left(f(p_{i,s_0^1}) + f(p_{i,s_0^2})\right) + \log_2(p_{i,s_0})\right)$$

$$= \frac{1}{4}\left(f(\sum_{i=0}^{3} p_{i,s_0}) - f(\sum_{i=0}^{3} p_{i,s_0^1}) - f(\sum_{i=0}^{3} p_{i,s_0^2})\right) - \frac{1}{4}\sum_{i=0}^{3}\left(f(p_{i,s_0}) - f(p_{i,s_0^1}) - f(p_{i,s_0^2})\right).$$

Note that the difference is only related to probabilities $p_{i,s_0}$, $p_{i,s_0^1}$, and $p_{i,s_0^2}$ for $i = 0, 1, 2, 3$.

To prove $I(X_1, X_2; \hat{Y}) \geqslant I(X_1, X_2; Y)$, we define a new function $g(u_1, u_2) = f(u_1 + u_2) - f(u_1) - f(u_2)$ and utilize its properties. We first prove $g(u_1 + v_1, u_2 + v_2) \geqslant g(u_1, u_2) + g(v_1, v_2)$ as follows:

$$g(u_1 + v_1, u_2 + v_2) - \left(g(u_1, u_2) + g(v_1, v_2)\right)$$

$$= (u_1 + u_2)\log_2(1 + \frac{v_1 + v_2}{u_1 + u_2}) + (v_1 + v_2)\log_2(1 + \frac{u_1 + u_2}{v_1 + v_2})$$

$$\quad - \left(u_1\log_2(1 + \frac{v_1}{u_1}) + v_1\log_2(1 + \frac{u_1}{v_1})\right) - \left(u_2\log_2(1 + \frac{v_2}{u_2}) + v_2\log_2(1 + \frac{u_2}{v_2})\right).$$

The function $t\log_2(1 + 1/t)$ is concave. Let $t_1 = \frac{u_1}{v_1}$, $t_2 = \frac{u_2}{v_2}$, $r_1 = \frac{v_1}{v_1 + v_2}$, and $r_2 = \frac{v_2}{v_1 + v_2}$. We have $(r_1 t_1 + r_2 t_2)\log_2\left(1 + 1/(r_1 t_1 + r_2 t_2)\right) \geqslant r_1 t_1\log_2(1 + 1/t_1) + r_2 t_2\log_2(1 + 1/t_2)$; that is, $(u_1 + u_2)\log_2(1 + \frac{v_1 + v_2}{u_1 + u_2}) \geqslant u_1\log_2(1 + \frac{v_1}{u_1}) + u_2\log_2(1 + \frac{v_2}{u_2})$. Similarly, $(v_1 + v_2)\log_2(1 + \frac{u_1 + u_2}{v_1 + v_2}) \geqslant v_1\log_2(1 + \frac{u_1}{v_1}) + v_2\log_2(1 + \frac{u_2}{v_2})$. Therefore, $g(u_1 + v_1, u_2 + v_2) \geqslant g(u_1, u_2) + g(v_1, v_2)$, where equality holds if and only if $\frac{u_1}{u_2} = \frac{v_1}{v_2}$.

Now, we apply $g(u_1 + v_1, u_2 + v_2) \geqslant g(u_1, u_2) + g(v_1, v_2)$ twice and have

$$f(\sum_{i=0}^{3} p_{i,s_0}) - f(\sum_{i=0}^{3} p_{i,s_0^1}) - f(\sum_{i=0}^{3} p_{i,s_0^2}) = g(p_{0,s_0^1} + p_{1,s_0^1} + p_{2,s_0^1} + p_{3,s_0^1}, p_{0,s_0^2} + p_{1,s_0^2} + p_{2,s_0^2} + p_{3,s_0^2})$$

$$\geqslant g(p_{0,s_0^1} + p_{1,s_0^1}, p_{0,s_0^2} + p_{1,s_0^2}) + g(p_{2,s_0^1} + p_{3,s_0^1}, p_{2,s_0^2} + p_{3,s_0^2})$$

$$\geqslant g(p_{0,s_0^1}, p_{0,s_0^2}) + g(p_{1,s_0^1}, p_{1,s_0^2}) + g(p_{2,s_0^1}, p_{2,s_0^2}) + g(p_{3,s_0^1}, p_{3,s_0^2})$$

$$= \sum_{i=0}^{3}\left(f(p_{i,s_0}) - f(p_{i,s_0^1}) - f(p_{i,s_0^2})\right),$$

where equality holds if and only if $\frac{p_{0,s_0^1}}{p_{0,s_0^2}} = \frac{p_{1,s_0^1}}{p_{1,s_0^2}} = \frac{p_{2,s_0^1}}{p_{2,s_0^2}} = \frac{p_{3,s_0^1}}{p_{3,s_0^2}}$. Thus, we have proved $I(X_1, X_2; \hat{Y}) \geqslant I(X_1, X_2; Y)$.

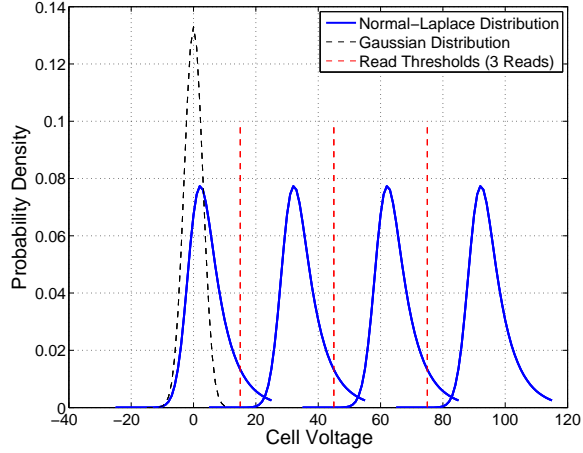

Fig. 5. Channel model for MLC flash memories with cell voltage modeled as the normal-Laplace distribution.

With the same proof technique, we can prove $I(X_1; \hat{Y}) \geqslant I(X_1; Y)$, $I(X_2; \hat{Y}) \geqslant I(X_2; Y)$, $I(X_1; \hat{Y}|X_2) \geqslant I(X_1; Y|X_2)$, and $I(X_2; \hat{Y}|X_1) \geqslant I(X_2; Y|X_1)$. These inequalities lead to $r_s^{TIN} \leqslant \hat{r}_s^{TIN}$, $\mathscr{R}^{TIN} \subseteq \hat{\mathscr{R}}^{TIN}$, and $\mathscr{R}^{SC} \subseteq \hat{\mathscr{R}}^{SC}$. ∎

In the following example, we show that the performance of MLC flash can be improved through multiple reads by simulations.

**Example 4** Following the results from [14], we assume the readback cell voltage has the normal-Laplace distribution $\mathcal{NL}(\mu, \nu, \alpha, \beta)$. The corresponding cumulative distribution function (cdf) for all real $y$ is

$$F(y) = \Phi\left(\frac{y - \mu}{\nu}\right) - \phi\left(\frac{y - \mu}{\nu}\right)\frac{\beta\mathfrak{R}(\alpha\nu - (y - \mu)/\nu) - \alpha\mathfrak{R}(\beta\nu + (y - \mu)/\nu)}{\alpha + \beta},$$

where $\Phi$ and $\phi$ are the cdf and probability density function (pdf) of a standard normal random variable and $\mathfrak{R}$ is Mills' ratio $\mathfrak{R}(z) = \frac{1 - \Phi(z)}{\phi(z)}$ [17]. As shown in Fig. 5, the readback cell voltage distributions for inputs $A_0$, $A_1$, $A_2$, and $A_3$ are $\mathcal{NL}(0, 3, 1/6, 1)$, $\mathcal{NL}(30, 3, 1/6, 1)$, $\mathcal{NL}(60, 3, 1/6, 1)$, and $\mathcal{NL}(90, 3, 1/6, 1)$, respectively. Compared to the Gaussian distribution, the normal-Laplace distribution captures the phenomenon of the cell voltage upward shift.

For the basic 3 reads setting, the read thresholds are placed at positions $\ell_a = 15$, $\ell_b = 45$, and $\ell_c = 75$. We use a vector to represent these positions $\vec{L}_3 = (15, 45, 75)$. Then, we apply additional reads to enhance the performance. Near each of positions $\ell_a$, $\ell_b$, and $\ell_c$, we add a few more reads. For example, near position $\ell_a$, we add a total of $2t$ additional reads at positions: $\ell_a - td, \ldots, \ell_a - 2d, \ell_a - d, \ell_a + d, \ell_a + 2d, \ldots, \ell_a + td$. For a total of 9 reads, with $t = 1$ and $d = 3$, the resulting read position vector is $\vec{L}_9 = (12, 15, 18, 42, 45, 48, 72, 75, 78)$. Similarly, for a total of 15 reads, with $t = 2$ and $d = 3$, the read position vector is $\vec{L}_{15} = (9, 12, 15, 18, 21, 39, 42, 45, 48, 51, 69, 72, 75, 78, 81)$. For a total of 21 reads, with $t = 3$ and $d = 3$, the read position vector is $\vec{L}_{21} = (6, 9, 12, 15, 18, 21, 24, 36, 39, 42, 45, 48, 51, 54, 66, 69, 72, 75,$

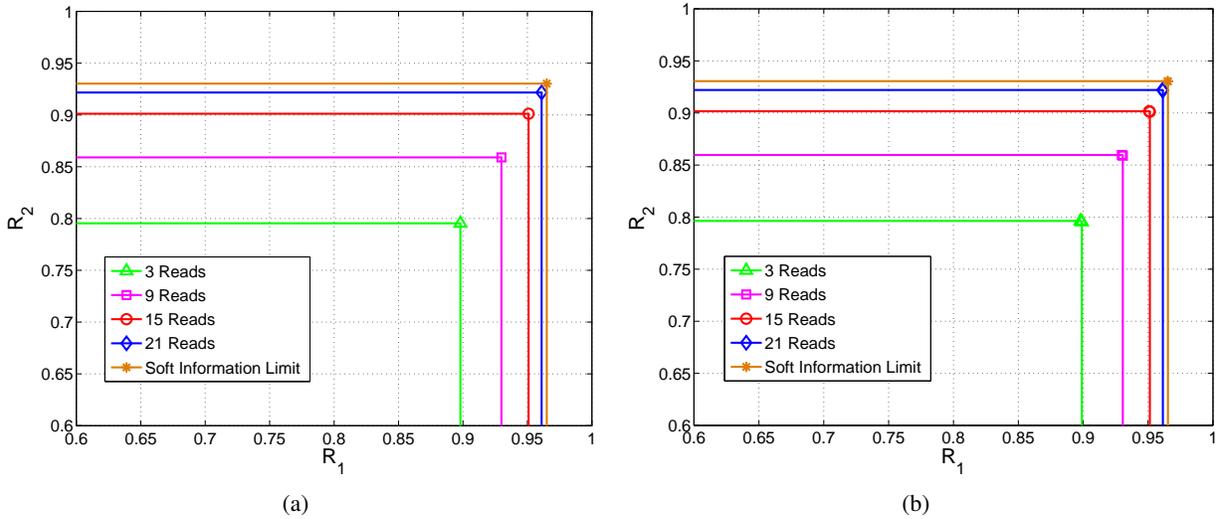

(a)                                                        (b)

Fig. 6.  Uniform rate regions under Gray labeling with different number of reads: (a) using TIN decoding, and (b) using SC decoding.

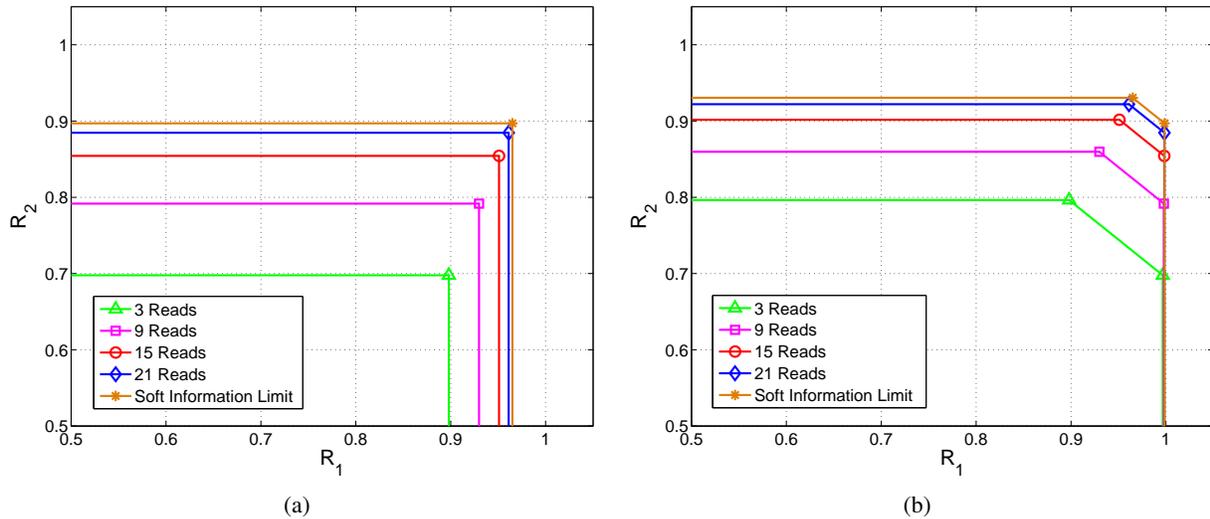

(a)                                                        (b)

Fig. 7.  Uniform rate regions under NO labeling with different number of reads: (a) using TIN decoding, and (b) using SC decoding.

78, 81, 84).

For Gray labeling, the uniform rate regions under TIN and SC decodings are plotted in Fig. 6. It is shown that the rate regions of TIN decoding and SC decoding are similar, due to Gray labeling. Additional reads can significantly improve the rates of both lower and upper pages. For NO labeling, the uniform rate regions under TIN and SC decodings are plotted in Fig. 7. Through additional reads, for either TIN decoding or SC decoding, the rate of the upper page is effectively enhanced. However, for SC decoding, the rate improvement of the lower page is very limited.

## V. Extension to TLC Flash Memory

In this section, we extend our analysis from MLC to TLC flash memory. Similar to the model proposed for MLC flash memory, given a labeling $\sigma$ and a quantizer $Q$, the TLC flash memory channel can be modeled as a 3-user discrete memoryless multiple access channel $\mathcal{W}_{TLC}$: $(\mathcal{X} \times \mathcal{X} \times \mathcal{X}, p(y|x_1, x_2, x_3), \mathcal{Y})$, where $\mathcal{X} = \{0, 1\}$, $\mathcal{Y} = \{s_0, s_1, \ldots, s_{q-1}\}$, and $p(y|x_1, x_2, x_3)$ is the transition probability for any $x_1, x_2, x_3 \in \mathcal{X}$ and $y \in \mathcal{Y}$. Define $P(Y = y|X_1 = x_1, X_2 = x_2, X_3 = x_3) \stackrel{\text{def}}{=} p_{BD(x_1, x_2, x_3), y}$, e.g., $P(Y = s_0 | X_1 = 1, X_2 = 0, X_3 = 1) = p_{5, s_0}$.

For a channel $\mathcal{W}_{TLC}$, for TIN decoding, its uniform rate region $\mathscr{R}^{TIN}$ is the set of all pairs $(R_1, R_2, R_3)$ such that: $0 \leqslant R_i \leqslant I(X_i; Y)$ for all $i = 1, 2, 3$. In $\mathscr{R}^{TIN}$, the sum rate $r_s^{TIN} = \max\{\sum_{i=1}^{3} R_i : (R_1, R_2, R_3) \in \mathscr{R}^{TIN}\} = \sum_{i=1}^{3} I(X_i, Y)$. For SC decoding, its uniform rate region $\mathscr{R}^{SC}$ is the set of all pairs $(R_1, R_2, R_3)$ such that:

1) $R_1 \leqslant I(X_1; Y | X_2, X_3)$, $R_2 \leqslant I(X_2; Y | X_1, X_3)$, $R_3 \leqslant I(X_3; Y | X_1, X_2)$;
2) $R_1 + R_2 \leqslant I(X_1, X_2; Y | X_3)$, $R_1 + R_3 \leqslant I(X_1, X_3; Y | X_2)$, $R_2 + R_3 \leqslant I(X_2, X_3; Y | X_1)$;
3) $R_1 + R_2 + R_3 \leqslant I(X_1, X_2, X_3; Y)$.

In $\mathscr{R}^{SC}$, the sum rate $r_s^{SC} = \max\{\sum_{i=1}^{3} R_i : (R_1, R_2, R_3) \in \mathscr{R}^{SC}\} = I(X_1, X_2, X_3; Y)$.

The following theorem gives the condition when the sum rates of TIN decoding and SC decoding are the same for a TLC flash channel.

**Theorem 15.** *For a channel $\mathcal{W}_{TLC}$, the sum rates $r_s^{TIN} \leqslant r_s^{SC}$ with equality if and only if*

$$
\begin{aligned}
&(p_{7,s_j} + p_{6,s_j})(p_{1,s_j} + p_{0,s_j}) = (p_{5,s_j} + p_{4,s_j})(p_{3,s_j} + p_{2,s_j}), \\
&p_{7,s_j}(p_{4,s_j} + p_{2,s_j} + p_{0,s_j}) = p_{6,s_j}(p_{5,s_j} + p_{3,s_j} + p_{1,s_j}), \\
&p_{5,s_j}(p_{6,s_j} + p_{2,s_j} + p_{0,s_j}) = p_{4,s_j}(p_{7,s_j} + p_{3,s_j} + p_{1,s_j}), \\
&p_{3,s_j}(p_{6,s_j} + p_{4,s_j} + p_{0,s_j}) = p_{2,s_j}(p_{7,s_j} + p_{5,s_j} + p_{1,s_j}), \\
&p_{1,s_j}(p_{6,s_j} + p_{4,s_j} + p_{2,s_j}) = p_{0,s_j}(p_{7,s_j} + p_{5,s_j} + p_{3,s_j}),
\end{aligned}
\tag{2}
$$

*for all $j = 0, 1, \ldots, q - 1$. If $r_s^{TIN} = r_s^{SC}$, then $\mathscr{R}^{TIN} = \mathscr{R}^{SC}$ and the rate region is a cube.*

*Proof:* We have $r_s^{SC} - r_s^{TIN}$ as follows:

$$
\begin{aligned}
r_s^{SC} - r_s^{TIN} = &I(X_1, X_2, X_3; Y) - \Big(I(X_1; Y) + I(X_2; Y) + I(X_3; Y)\Big) \\
= &I(X_1; Y) + I(X_2; Y | X_1) + I(X_3; Y | X_1, X_2) - \Big(I(X_1; Y) + I(X_2; Y) + I(X_3; Y)\Big) \\
= &H(X_2 | X_1) - H(X_2 | Y, X_1) + H(X_3 | X_1, X_2) - H(X_3 | Y, X_1, X_2) \\
&- \Big(H(X_2) - H(X_2 | Y) + H(X_3) - H(X_3 | Y)\Big) \\
\overset{(a)}{=} &I(X_1; X_2 | Y) + I(X_1, X_2; X_3 | Y) \\
= &\sum_{j=0}^{q-1} (\sum_{i=0}^{7} \frac{p_{i,s_j}}{8}) I(X_1; X_2 | Y = s_j) + \sum_{j=0}^{q-1} (\sum_{i=0}^{7} \frac{p_{i,s_j}}{8}) I(X_1, X_2; X_3 | Y = s_j) \geqslant 0,
\end{aligned}
$$

where step $(a)$ is due to the fact that $X_1$, $X_2$, and $X_3$ are independent. The last step is from that $I(X_1; X_2|Y = s_j) \geqslant 0$ and $I(X_1, X_2; X_3|Y = s_j) \geqslant 0$ for all $j = 0, 1, \ldots, q-1$. Therefore, $r_s^{TIN} = r_s^{SC}$ if and only if $I(X_1; X_2|Y = s_j) = 0$ and $I(X_1, X_2; X_3|Y = s_j) = 0$ for all $j = 0, 1, \ldots, q-1$.

The condition $I(X_1; X_2|Y = s_j) = 0$ will be satisfied if and only if $X_1$ and $X_2$ are conditionally independent given $Y = s_j$, i.e., $P(X_1, X_2|Y = s_j) = P(X_1|Y = s_j)P(X_2|Y = s_j)$. Similar to the proof of Theorem 1, the condition $P(X_1, X_2|Y = s_j) = P(X_1|Y = s_j)P(X_2|Y = s_j)$ leads to the first condition in the statement of the theorem.

Similarly, $I(X_1, X_2; X_3|Y = s_j) = 0$ will be satisfied if and only if $(X_1, X_2)$ and $X_3$ are conditionally independent given $Y = s_j$, i.e., $P(X_1, X_2, X_3/Y = s_j) = P(X_1, X_2/Y = s_j)P(X_3/Y = s_j)$, which results in the last four conditions in the statement of the theorem.

Next, we prove $\mathscr{R}^{TIN} = \mathscr{R}^{SC}$ if $r_s^{TIN} = r_s^{SC}$. Assuming $r_s^{TIN} = r_s^{SC}$, i.e., $I(X_1, X_2, X_3; Y) = I(X_1; Y) + I(X_2; Y) + I(X_3; Y)$, from above we have already shown that $I(X_1; X_2|Y) = 0$ and $I(X_1, X_2; X_3|Y) = 0$. Similarly, we have $I(X_1; X_3|Y) = 0$, $I(X_1, X_3; X_2|Y) = 0$, $I(X_2; X_3|Y) = 0$, and $I(X_2, X_3; X_1|Y) = 0$. To prove $\mathscr{R}^{TIN} = \mathscr{R}^{SC}$, we need to show: 1) $I(X_1; Y|X_2, X_3) = I(X_1; Y)$, 2) $I(X_2; Y|X_1, X_3) = I(X_2; Y)$, 3) $I(X_3; Y|X_1, X_2) = I(X_3; Y)$, 4) $I(X_1, X_2; Y|X_3) = I(X_1; Y) + I(X_2; Y)$, 5) $I(X_1, X_3; Y|X_2) = I(X_1; Y) + I(X_3; Y)$, 6) $I(X_2, X_3; Y|X_1) = I(X_2; Y) + I(X_3; Y)$.

In the following, we will prove two of the above six equations. They are $I(X_3; Y|X_1, X_2) = I(X_3; Y)$ and $I(X_1, X_2; Y|X_3) = I(X_1; Y) + I(X_2; Y)$. The other four equations can be proved in a similar way. For the first, we have

$$
\begin{aligned}
I(X_3; Y|X_1, X_2) - I(X_3; Y) =& H(X_3|X_1, X_2) - H(X_3|Y, X_1, X_2) - \Big(H(X_3) - H(X_3|Y)\Big) \\
\overset{(a)}{=}& H(X_3|Y) - H(X_3|Y, X_1, X_2) = I(X_3; X_1, X_2|Y) = 0,
\end{aligned}
$$

where step $(a)$ is due to the fact that $X_1$, $X_2$, and $X_3$ are independent. Similarly, for the second,

$$
\begin{aligned}
&I(X_1, X_2; Y|X_3) - I(X_1; Y) - I(X_2; Y) \\
=& H(X_1, X_2|X_3) - H(X_1, X_2|Y, X_3) - \Big(H(X_1) - H(X_1|Y)\Big) - \Big(H(X_2) - H(X_2|Y)\Big) \\
\overset{(a)}{=}& H(X_1|Y) + H(X_2|Y) - H(X_1, X_2|Y, X_3) \\
\overset{(b)}{=}& H(X_1|Y) + H(X_2|Y) - H(X_1, X_2|Y) = I(X_1; X_2|Y) = 0,
\end{aligned}
$$

where step $(a)$ is due to the fact that $X_1$, $X_2$, and $X_3$ are independent, and step (b) is from $I(X_1, X_2; X_3|Y) = H(X_1, X_2|Y) - H(X_1, X_2|Y, X_3) = 0$. ∎

The upper bound on the difference between the sum rates $r_s^{SC}$ and $r_s^{TIN}$ is given by the following theorem.

**Theorem 16.** *For a channel $\mathcal{W}_{TLC}$, the rate difference $r_s^{SC} - r_s^{TIN} \leqslant 2$ with equality if and only if*

$$
\begin{aligned}
&p_{7,s_j} + p_{6,s_j} + p_{5,s_j} + p_{4,s_j} = p_{3,s_j} + p_{2,s_j} + p_{1,s_j} + p_{0,s_j}, \\
&p_{7,s_j} + p_{5,s_j} + p_{3,s_j} + p_{1,s_j} = p_{6,s_j} + p_{4,s_j} + p_{2,s_j} + p_{0,s_j}, \\
&(p_{7,s_j} + p_{6,s_j})(p_{3,s_j} + p_{2,s_j}) = (p_{5,s_j} + p_{4,s_j})(p_{1,s_j} + p_{0,s_j}) = 0, \\
&p_{7,s_j}p_{6,s_j} = p_{5,s_j}p_{4,s_j} = p_{3,s_j}p_{2,s_j} = p_{1,s_j}p_{0,s_j} = 0,
\end{aligned}
\tag{3}
$$

*for all $j = 0, 1, \ldots, q - 1$.*

*Proof:* We bound $r_s^{SC} - r_s^{TIN}$ as

$$
\begin{aligned}
r_s^{SC} - r_s^{TIN} &= I(X_1; X_2 | Y) + I(X_1, X_2; X_3 | Y) = H(X_1 | Y) - H(X_1 | X_2, Y) + H(X_3 | Y) - H(X_3 | X_1, X_2, Y) \\
&\overset{(a)}{\leqslant} H(X_1) + H(X_3) - H(X_1 | X_2, Y) - H(X_3 | X_1, X_2, Y) \overset{(b)}{\leqslant} H(X_1) + H(X_3) = 2,
\end{aligned}
$$

where step (a) follows from $H(X_1 | Y) \leqslant H(X_1)$ and $H(X_3 | Y) \leqslant H(X_3)$, and step (b) follows from $H(X_1 | X_2, Y) \geqslant 0$ and $H(X_3 | X_1, X_2, Y) \geqslant 0$. Thus, $r_s^{SC} - r_s^{TIN} = 2$ if and only if 1) $H(X_1 | Y) = H(X_1) = H(X_3 | Y) = H(X_3) = 1$, and 2) $H(X_1 | X_2, Y) = H(X_3 | X_1, X_2, Y) = 0$.

The condition $H(X_1 | Y) = \sum_{j=0}^{q-1} (\sum_{i=0}^{7} \frac{p_{i,s_j}}{8}) H(X_1 | Y = s_j) = 1$ holds if and only if $p_{7,s_j} + p_{6,s_j} + p_{5,s_j} + p_{4,s_j} = p_{3,s_j} + p_{2,s_j} + p_{1,s_j} + p_{0,s_j}$, for all $j = 0, 1, \ldots, q - 1$.

Similarly, $H(X_3 | Y) = 1$ gives that $p_{7,s_j} + p_{5,s_j} + p_{3,s_j} + p_{1,s_j} = p_{6,s_j} + p_{4,s_j} + p_{2,s_j} + p_{0,s_j}$, $H(X_1 | X_2, Y) = 0$ results in $(p_{7,s_j} + p_{6,s_j})(p_{3,s_j} + p_{2,s_j}) = (p_{5,s_j} + p_{4,s_j})(p_{1,s_j} + p_{0,s_j}) = 0$, and $H(X_3 | X_1, X_2, Y) = 0$ requires that $p_{7,s_j}p_{6,s_j} = p_{5,s_j}p_{4,s_j} = p_{3,s_j}p_{2,s_j} = p_{1,s_j}p_{0,s_j} = 0$, for all $j = 0, 1, \ldots, q - 1$. ∎

**Definition 17** *For TLC flash memory, labeling $\sigma_G$=(111, 110, 100, 101, 001, 000, 010, 011) is called Gray labeling, and $\sigma_{NO}$=(111, 110, 101, 100, 011, 010, 001, 000) is called Natural Order (NO) labeling.*

For each labeling, the mapping between inputs $(X_1, X_2, X_3) \in \mathcal{T}_{TLC}$ and voltage levels $V \in \mathcal{V}_{TLC}$ is shown in Table VI. We fix our quantizer $Q$ with seven reads, which are placed between every two neighboring voltage levels, as shown in Fig. 1(b). Hence, the output alphabet is $\mathcal{Y}_{TLC} = \{s_0, s_1, \ldots, s_7\}$.

We consider a simple P/E cycling model, called the *early stage P/E cycling* model, characterized by the channel transition matrix $p_{TLC}^E(y|v)$, for $y \in \mathcal{Y}_{TLC}$ and $v \in \mathcal{V}_{TLC}$, shown in Table VI, where $\varepsilon_i$ and $1 - \varepsilon_i$, $i = 0, 1, \ldots, 6$, represent non-zero transition probabilities. This model reflects the phenomenon of upward shift of the cell voltage. A related model is the *data retention* model, where the channel transition matrix $p_{TLC}^{DR}(y|v)$, for output $y \in \mathcal{Y}_{TLC}$ and voltage level $v \in \mathcal{V}_{TLC}$, reflects the downward shift of cell voltage and its structure is shown in Table VII, where $\epsilon_i$ and $1 - \epsilon_i$, $i = 1, 2, \ldots, 7$, represent non-zero transition probabilities. We give an analysis of the early stage P/E cycling model below. This is meaningful, since the techniques and results apply to the data retention model as well. For a more accurate P/E cycling model, further experimental studies are required. Once the channel model is obtained, however, similar analysis can be carried out.

TABLE VI
Channel transition matrix $p_{TLC}^E(y|v)$ at early stage of $P/E$ cycling for TLC flash memories

| $V$ | Inputs: $(X_1, X_2, X_3)$ | | | | | Output: $Y$ | | | | |
|---|---|---|---|---|---|---|---|---|---|---|
| Levels | Gray | NO | $s_0$ | $s_1$ | $s_2$ | $s_3$ | $s_4$ | $s_5$ | $s_6$ | $s_7$ |
| $B_0$ | (111) | (111) | $\varepsilon_0$ | 1- $\varepsilon_0$ | 0 | 0 | 0 | 0 | 0 | 0 |
| $B_1$ | (110) | (110) | 0 | $\varepsilon_1$ | $1 - \varepsilon_1$ | 0 | 0 | 0 | 0 | 0 |
| $B_2$ | (100) | (101) | 0 | 0 | $\varepsilon_2$ | $1 - \varepsilon_2$ | 0 | 0 | 0 | 0 |
| $B_3$ | (101) | (100) | 0 | 0 | 0 | $\varepsilon_3$ | $1 - \varepsilon_3$ | 0 | 0 | 0 |
| $B_4$ | (001) | (011) | 0 | 0 | 0 | 0 | $\varepsilon_4$ | 1-$\varepsilon_4$ | 0 | 0 |
| $B_5$ | (000) | (010) | 0 | 0 | 0 | 0 | 0 | $\varepsilon_5$ | 1-$\varepsilon_5$ | 0 |
| $B_6$ | (010) | (001) | 0 | 0 | 0 | 0 | 0 | 0 | $\varepsilon_6$ | 1-$\varepsilon_6$ |
| $B_7$ | (011) | (000) | 0 | 0 | 0 | 0 | 0 | 0 | 0 | 1 |

TABLE VII
Channel transition matrix $p_{TLC}^{DR}(y|v)$ of data retention model for TLC flash memories

| $V$ | Inputs: $(X_1, X_2, X_3)$ | | | | | Output: $Y$ | | | | |
|---|---|---|---|---|---|---|---|---|---|---|
| Levels | Gray | NO | $s_0$ | $s_1$ | $s_2$ | $s_3$ | $s_4$ | $s_5$ | $s_6$ | $s_7$ |
| $B_0$ | (111) | (111) | 1 | 0 | 0 | 0 | 0 | 0 | 0 | 0 |
| $B_1$ | (110) | (110) | $1 - \varepsilon_1$ | $\varepsilon_1$ | 0 | 0 | 0 | 0 | 0 | 0 |
| $B_2$ | (100) | (101) | 0 | $1 - \varepsilon_2$ | $\varepsilon_2$ | 0 | 0 | 0 | 0 | 0 |
| $B_3$ | (101) | (100) | 0 | 0 | $1 - \varepsilon_3$ | $\varepsilon_3$ | 0 | 0 | 0 | 0 |
| $B_4$ | (001) | (011) | 0 | 0 | 0 | $1 - \varepsilon_4$ | $\varepsilon_4$ | 0 | 0 | 0 |
| $B_5$ | (000) | (010) | 0 | 0 | 0 | 0 | $1 - \varepsilon_5$ | $\varepsilon_5$ | 0 | 0 |
| $B_6$ | (010) | (001) | 0 | 0 | 0 | 0 | 0 | $1 - \varepsilon_6$ | $\varepsilon_6$ | 0 |
| $B_7$ | (011) | (000) | 0 | 0 | 0 | 0 | 0 | 0 | $1 - \varepsilon_7$ | $\varepsilon_7$ |

For the early stage $P/E$ cycling model, we have the following lemma.

**Lemma 18.** *For channel transition matrix $p_{TLC}^E(y|v)$, using Gray labeling, we have $r_s^{TIN} = r_s^{SC}$ and $\mathscr{R}^{TIN} = \mathscr{R}^{SC}$. Using NO labeling, we have $r_s^{TIN} < r_s^{SC}$.*

*Proof:* With Gray labeling, we can verify that the conditions, i.e., the set of equations (2), in Theorem 15, are satisfied for $j = 0, 1, \ldots, 7$. Thus, from Theorem 15, we conclude $r_s^{TIN} = r_s^{SC}$ and $\mathscr{R}^{TIN} = \mathscr{R}^{SC}$. While under NO labeling $p_{7,s_2}(p_{4,s_2} + p_{2,s_2} + p_{0,s_2}) \neq p_{6,s_2}(p_{5,s_2} + p_{3,s_2} + p_{1,s_2})$. Thus, in this case, $r_s^{TIN} < r_s^{SC}$. ∎

Next, for the early stage of P/E cycling model, we make the following observations about the corner points of uniform rate regions for TIN and SC decodings under Gray and NO labelings: 1) with Gray labeling, from Theorem 15 and Lemma 18, for both TIN and SC decoding, we only have one same corner point: $\mathbb{P}_G^0 = \left( I(X_1; Y), I(X_2; Y), I(X_3; Y) \right)$; 2) with NO labeling, for TIN decoding, the corner point is: $\mathbb{P}_{NO}^0 = \left( I(X_1; Y), I(X_2; Y), I(X_3; Y) \right)$ [4]. For SC decoding, we have six corner points: $\mathbb{P}_{NO}^1 = \left( I(X_1; Y), I(X_2; Y|X_1), I(X_3; Y|X_1, X_2) \right)$, $\mathbb{P}_{NO}^2 = \left( I(X_1; Y), I(X_2; Y|X_1, X_3), I(X_3; Y|X_1) \right)$, $\mathbb{P}_{NO}^3 = \left( I(X_1; Y|X_2), I(X_2; Y), I(X_3; Y|X_1, X_2) \right)$, $\mathbb{P}_{NO}^4 = \left( I(X_1; Y|X_2, X_3), I(X_2; Y), I(X_3; Y|X_2) \right)$, $\mathbb{P}_{NO}^5 = \left( I(X_1; Y|X_3), I(X_2; Y|X_1, X_3), I(X_3; Y) \right)$,

---

[4] Although there is a term $I(X_1; Y)$ in both $\mathbb{P}_G^0$ and $\mathbb{P}_{NO}^0$, they represent different values due to the use of different labelings for the two points $\mathbb{P}_G^0$ and $\mathbb{P}_{NO}^0$. The same applies to other common terms.

$\mathbb{P}_{NO}^6 = \big(I(X_1;Y|X_2,X_3),\ I(X_2;Y|X_3),\ I(X_3;Y)\big)$. We have the following structure of these corner points.

**Theorem 19.** *With channel transition matrix $p_{TLC}^E(y|v)$, the corner points satisfy $\mathbb{P}_G^0 = \mathbb{P}_{NO}^1$. For SC decoding, we have point $\mathbb{P}_{NO}^2 = \big(I(X_1;Y),1,I(X_3;Y|X_1)\big)$, points $\mathbb{P}_{NO}^3 = \mathbb{P}_{NO}^4 = \big(1,I(X_2;Y),I(X_3;Y|X_1,X_2)\big)$, and points $\mathbb{P}_{NO}^5 = \mathbb{P}_{NO}^6 = \big(1,1,I(X_3;Y)\big)$.*

*Proof:* To prove $\mathbb{P}_G^0 = \mathbb{P}_{NO}^1$, we first show the first coordinate $\mathbb{P}_G^0(1)$ of point $\mathbb{P}_G^0$ is equal to the first coordinate $\mathbb{P}_{NO}^1(1)$ of point $\mathbb{P}_{NO}^1$.

$\mathbb{P}_G^0(1)$

$=I(X_1;Y) = H(Y) - H(Y|X_1)$

$=H(\frac{\varepsilon_0}{8}, \frac{1-\varepsilon_0+\varepsilon_1}{8}, \frac{1-\varepsilon_1+\varepsilon_2}{8}, \frac{1-\varepsilon_2+\varepsilon_3}{8}, \frac{1-\varepsilon_3+\varepsilon_4}{8}, \frac{1-\varepsilon_4+\varepsilon_5}{8}, \frac{1-\varepsilon_5+\varepsilon_6}{8}, \frac{1-\varepsilon_6+1}{8})$

$\quad -\frac{1}{2}H(\frac{\varepsilon_0}{4}, \frac{1-\varepsilon_0+\varepsilon_1}{4}, \frac{1-\varepsilon_1+\varepsilon_2}{4}, \frac{1-\varepsilon_2+\varepsilon_3}{4}, \frac{1-\varepsilon_3}{4}) - \frac{1}{2}H(\frac{\varepsilon_4}{4}, \frac{1-\varepsilon_4+\varepsilon_5}{4}, \frac{1-\varepsilon_5+\varepsilon_6}{4}, \frac{1-\varepsilon_6+1}{4})$

$=\mathbb{P}_{NO}^1(1)$.

Now, we show the second coordinate $\mathbb{P}_G^0(2)$ of point $\mathbb{P}_G^0$ equals the second coordinate $\mathbb{P}_{NO}^1(2)$ of point $\mathbb{P}_{NO}^1$.

$\mathbb{P}_G^0(2) = I(X_2;Y) = H(Y) - H(Y|X_2)$

$\quad =H(\frac{\varepsilon_0}{8}, \frac{1-\varepsilon_0+\varepsilon_1}{8}, \frac{1-\varepsilon_1+\varepsilon_2}{8}, \frac{1-\varepsilon_2+\varepsilon_3}{8}, \frac{1-\varepsilon_3+\varepsilon_4}{8}, \frac{1-\varepsilon_4+\varepsilon_5}{8}, \frac{1-\varepsilon_5+\varepsilon_6}{8}, \frac{1-\varepsilon_6+1}{8})$

$\quad -\frac{1}{2}H(\frac{\varepsilon_0}{4}, \frac{1-\varepsilon_0+\varepsilon_1}{4}, \frac{1-\varepsilon_1}{4}, \frac{\varepsilon_6}{4}, \frac{1-\varepsilon_6+1}{4}) - \frac{1}{2}H(\frac{\varepsilon_2}{4}, \frac{1-\varepsilon_2+\varepsilon_3}{4}, \frac{1-\varepsilon_3+\varepsilon_4}{4}, \frac{1-\varepsilon_4+\varepsilon_5}{4}, \frac{1-\varepsilon_5}{4})$

$\quad \overset{(a)}{=} H(\frac{\varepsilon_0}{8}, \frac{1-\varepsilon_0+\varepsilon_1}{8}, \frac{1-\varepsilon_1+\varepsilon_2}{8}, \frac{1-\varepsilon_2+\varepsilon_3}{8}, \frac{1-\varepsilon_3}{8}, \frac{\varepsilon_4}{8}, \frac{1-\varepsilon_4+\varepsilon_5}{8}, \frac{1-\varepsilon_5+\varepsilon_6}{8}, \frac{1-\varepsilon_6+1}{8})$

$\quad -\frac{1-\varepsilon_3+\varepsilon_4}{8}H(\frac{1-\varepsilon_3}{1-\varepsilon_3+\varepsilon_4}, \frac{\varepsilon_4}{1-\varepsilon_3+\varepsilon_4}) - \frac{1}{2}H(\frac{\varepsilon_0}{4}, \frac{1-\varepsilon_0+\varepsilon_1}{4}, \frac{1-\varepsilon_1}{4}, \frac{\varepsilon_6}{4}, \frac{1-\varepsilon_6+1}{4})$

$\quad -\frac{1}{2}\Big(H(\frac{\varepsilon_2}{4}, \frac{1-\varepsilon_2+\varepsilon_3}{4}, \frac{1-\varepsilon_3}{4}, \frac{\varepsilon_4}{4}, \frac{1-\varepsilon_4+\varepsilon_5}{4}, \frac{1-\varepsilon_5}{4}) - \frac{1-\varepsilon_3+\varepsilon_4}{4}H(\frac{1-\varepsilon_3}{1-\varepsilon_3+\varepsilon_4}, \frac{\varepsilon_4}{1-\varepsilon_3+\varepsilon_4})\Big)$

$\quad \overset{(b)}{=} H(\frac{1}{2},\frac{1}{2}) + \frac{1}{2}H(\frac{\varepsilon_0}{4}, \frac{1-\varepsilon_0+\varepsilon_1}{4}, \frac{1-\varepsilon_1+\varepsilon_2}{4}, \frac{1-\varepsilon_2+\varepsilon_3}{4}, \frac{1-\varepsilon_3}{4}) + \frac{1}{2}H(\frac{\varepsilon_4}{4}, \frac{1-\varepsilon_4+\varepsilon_5}{4}, \frac{1-\varepsilon_5+\varepsilon_6}{4}, \frac{2-\varepsilon_6}{4})$

$\quad -\frac{1}{2}\Big(H(\frac{1}{2},\frac{1}{2}) + \frac{1}{2}H(\frac{\varepsilon_0}{2}, \frac{1-\varepsilon_0+\varepsilon_1}{2}, \frac{1-\varepsilon_1}{2}) + \frac{1}{2}H(\frac{\varepsilon_6}{2}, \frac{2-\varepsilon_6}{2})\Big)$

$\quad -\frac{1}{2}\Big(H(\frac{1}{2},\frac{1}{2}) + \frac{1}{2}H(\frac{\varepsilon_2}{2}, \frac{1-\varepsilon_2+\varepsilon_3}{2}, \frac{1-\varepsilon_3}{2}) + \frac{1}{2}H(\frac{\varepsilon_4}{2}, \frac{1-\varepsilon_4+\varepsilon_5}{2}, \frac{1-\varepsilon_5}{2})\Big)$

$\quad =\frac{1}{2}H(\frac{\varepsilon_0}{4}, \frac{1-\varepsilon_0+\varepsilon_1}{4}, \frac{1-\varepsilon_1+\varepsilon_2}{4}, \frac{1-\varepsilon_2+\varepsilon_3}{4}, \frac{1-\varepsilon_3}{4}) + \frac{1}{2}H(\frac{\varepsilon_4}{4}, \frac{1-\varepsilon_4+\varepsilon_5}{4}, \frac{1-\varepsilon_5+\varepsilon_6}{4}, \frac{2-\varepsilon_6}{4})$

$\quad -\frac{1}{4}\Big(H(\frac{\varepsilon_0}{2}, \frac{1-\varepsilon_0+\varepsilon_1}{2}, \frac{1-\varepsilon_1}{2}) + H(\frac{\varepsilon_2}{2}, \frac{1-\varepsilon_2+\varepsilon_3}{2}, \frac{1-\varepsilon_3}{2}) + H(\frac{\varepsilon_4}{2}, \frac{1-\varepsilon_4+\varepsilon_5}{2}, \frac{1-\varepsilon_5}{2}) + H(\frac{\varepsilon_6}{2}, \frac{2-\varepsilon_6}{2})\Big)$

$\quad =\mathbb{P}_{NO}^1(2),$

where steps (a) and (b) follow from the grouping property of entropy [15].

Since the sum of the three coordinates of point $\mathbb{P}_G^0$ is the same as that of point $\mathbb{P}_{NO}^1$, the third coordinate of point $\mathbb{P}_G^0$ is also the same as that of point $\mathbb{P}_{NO}^1$. Thus, we have proved $\mathbb{P}_G^0 = \mathbb{P}_{NO}^1$.

Next, the second coordinate $\mathbb{P}_{NO}^2(2)$ of point $\mathbb{P}_{NO}^2$ is 1, since $\mathbb{P}_{NO}^2(2) = I(X_2;Y|X_1,X_3) =$

$H(X_2|X_1,X_3) - H(X_2|X_1,X_3,Y) = H(X_2) = 1$.

For point $\mathbb{P}_{NO}^3$, its first coordinate $\mathbb{P}_{NO}^3(1) = I(X_1;Y|X_2) = H(X_1|X_2) - H(X_1|X_2,Y) = H(X_1) = 1$. For point $\mathbb{P}_{NO}^4$, its first coordinate $\mathbb{P}_{NO}^4(1) = I(X_1;Y|X_2,X_3) = H(X_1|X_2,X_3) - H(X_1|X_2,X_3,Y) = H(X_1) = 1$. Thus, the first and second coordinates of points $\mathbb{P}_{NO}^3$ and $\mathbb{P}_{NO}^4$ are the same. Since the sum of the three coordinates of point $\mathbb{P}_{NO}^3$ is the same as that of point $\mathbb{P}_{NO}^4$, the third coordinate of point $\mathbb{P}_{NO}^3$ is also equal to that of point $\mathbb{P}_{NO}^4$.

Similarly, to prove $\mathbb{P}_{NO}^5 = \mathbb{P}_{NO}^6$, we only need to show their first two coordinates are the same. For point $\mathbb{P}_{NO}^5$, its first coordinate $\mathbb{P}_{NO}^5(1) = I(X_1;Y|X_3) = H(X_1|X_3) - H(X_1|X_3,Y) = H(X_1) = 1$, and its second coordinate $\mathbb{P}_{NO}^5(2) = \mathbb{P}_{NO}^2(2) = 1$. For point $\mathbb{P}_{NO}^6$, its first coordinate $\mathbb{P}_{NO}^6(1) = \mathbb{P}_{NO}^4(1) = 1$ and its second coordinate $\mathbb{P}_{NO}^6(2) = I(X_2;Y|X_3) = H(X_2|X_3) - H(X_2|X_3,Y) = H(X_2) = 1$. ■

Let the uniform rate regions of TIN decoding and SC decoding under Gray labeling be $\mathscr{R}_G^{TIN}$ and $\mathscr{R}_G^{SC}$, respectively, and under NO labeling be $\mathscr{R}_{NO}^{TIN}$ and $\mathscr{R}_{NO}^{SC}$, respectively. We have the following theorem on the uniform rate regions.

**Theorem 20.** *With channel transition matrix $p_{TLC}^E(y|v)$, the uniform rate regions satisfy $\mathscr{R}_{NO}^{TIN} \subset \mathscr{R}_G^{TIN} = \mathscr{R}_G^{SC} \subset \mathscr{R}_{NO}^{SC}$.*

*Proof:* From Lemma 18, we have $\mathscr{R}_G^{TIN} = \mathscr{R}_G^{SC}$. From Theorem 19, it is clear to see $\mathscr{R}_G^{SC} \subset \mathscr{R}_{NO}^{SC}$. Here, we only need to show $\mathscr{R}_{NO}^{TIN} \subset \mathscr{R}_G^{TIN}$. We prove it by showing that $\mathbb{P}_G^0(1) = \mathbb{P}_{NO}^0(1)$, $\mathbb{P}_G^0(2) > \mathbb{P}_{NO}^0(2)$, and $\mathbb{P}_G^0(3) > \mathbb{P}_{NO}^0(3)$. First, from Theorem 19, $\mathbb{P}_G^0(1) = \mathbb{P}_{NO}^1(1)$. Since $\mathbb{P}_{NO}^0(1) = \mathbb{P}_{NO}^1(1)$, we have $\mathbb{P}_G^0(1) = \mathbb{P}_{NO}^0(1)$. Next, we show $\mathbb{P}_G^0(2) > \mathbb{P}_{NO}^0(2)$ as follows:

$$\mathbb{P}_G^0(2) - \mathbb{P}_{NO}^0(2)$$
$$= \frac{1}{2}H(\frac{\varepsilon_0}{4}, \frac{1-\varepsilon_0+\varepsilon_1}{4}, \frac{1-\varepsilon_1}{4}, \frac{\varepsilon_4}{4}, \frac{1-\varepsilon_4+\varepsilon_5}{4}, \frac{1-\varepsilon_5}{4}) + \frac{1}{2}H(\frac{\varepsilon_2}{4}, \frac{1-\varepsilon_2+\varepsilon_3}{4}, \frac{1-\varepsilon_3}{4}, \frac{\varepsilon_6}{4}, \frac{1-\varepsilon_6+1}{4})$$
$$- \frac{1}{2}H(\frac{\varepsilon_0}{4}, \frac{1-\varepsilon_0+\varepsilon_1}{4}, \frac{1-\varepsilon_1}{4}, \frac{\varepsilon_6}{4}, \frac{1-\varepsilon_6+1}{4}) - \frac{1}{2}H(\frac{\varepsilon_2}{4}, \frac{1-\varepsilon_2+\varepsilon_3}{4}, \frac{1-\varepsilon_3+\varepsilon_4}{4}, \frac{1-\varepsilon_4+\varepsilon_5}{4}, \frac{1-\varepsilon_5}{4})$$
$$= \frac{1}{8}\Big(f(1-\varepsilon_3+\varepsilon_4) - f(1-\varepsilon_3) - f(\varepsilon_4)\Big) = \frac{1}{8}\Big((1-\varepsilon_3)\log_2(\frac{1-\varepsilon_3+\varepsilon_4}{1-\varepsilon_3}) + \varepsilon_4\log_2(\frac{1-\varepsilon_3+\varepsilon_4}{\varepsilon_4})\Big) > 0.$$

Similarly, we show $\mathbb{P}_G^0(3) > \mathbb{P}_{NO}^0(3)$:

$$\mathbb{P}_G^0(3) - \mathbb{P}_{NO}^0(3)$$
$$= \frac{1}{2}H(\frac{\varepsilon_0}{4}, \frac{1-\varepsilon_0}{4}, \frac{\varepsilon_2}{4}, \frac{1-\varepsilon_2}{4}, \frac{\varepsilon_4}{4}, \frac{1-\varepsilon_4}{4}, \frac{\varepsilon_6}{4}, \frac{1-\varepsilon_6}{4}) + \frac{1}{2}H(\frac{\varepsilon_1}{4}, \frac{1-\varepsilon_1}{4}, \frac{\varepsilon_3}{4}, \frac{1-\varepsilon_3}{4}, \frac{\varepsilon_5}{4}, \frac{1-\varepsilon_5}{4}, \frac{1}{4})$$
$$- \frac{1}{2}H(\frac{\varepsilon_0}{4}, \frac{1-\varepsilon_0}{4}, \frac{\varepsilon_3}{4}, \frac{1-\varepsilon_3+\varepsilon_4}{4}, \frac{1-\varepsilon_4}{4}, \frac{1}{4}) - \frac{1}{2}H(\frac{\varepsilon_1}{4}, \frac{1-\varepsilon_1+\varepsilon_2}{4}, \frac{1-\varepsilon_2}{4}, \frac{\varepsilon_5}{4}, \frac{1-\varepsilon_5+\varepsilon_6}{4}, \frac{1-\varepsilon_6}{4})$$
$$= \frac{1}{8}\Big(f(1-\varepsilon_1+\varepsilon_2) - f(1-\varepsilon_1) - f(\varepsilon_2)\Big) + \frac{1}{8}\Big(f(1-\varepsilon_3+\varepsilon_4) - f(1-\varepsilon_3) - f(\varepsilon_4)\Big)$$
$$+ \frac{1}{8}\Big(f(1-\varepsilon_5+\varepsilon_6) - f(1-\varepsilon_5) - f(\varepsilon_6)\Big) > 0.$$

■

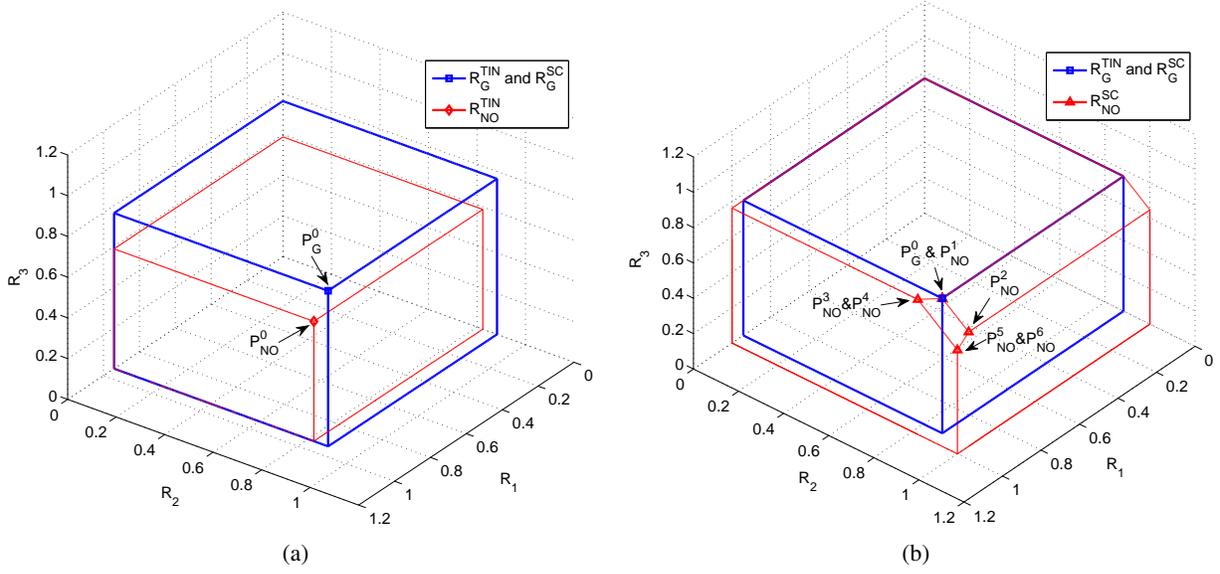

Fig. 8. (a) Illustration of uniform rate regions for TIN decoding under Gray labeling $\mathscr{R}_G^{TIN}$, SC decoding under Gray labeling $\mathscr{R}_G^{SC}$, and TIN decoding under NO labeling $\mathscr{R}_{NO}^{TIN}$. (b) Illustration of uniform rate regions for TIN decoding under Gray labeling $\mathscr{R}_G^{TIN}$, SC decoding under Gray labeling $\mathscr{R}_G^{SC}$, and SC decoding under NO labeling $\mathscr{R}_{NO}^{SC}$.

**Remark 4** Similar to the MLC flash case, in TLC flash, for the early stage P/E cycling model, from Lemma 18 and Theorems 19 and 20, we have the following observations: 1) The sum rate of TIN decoding under Gray labeling is the same as that of SC decoding. This implies that an ECC solution for Gray labeling need only use three point-to-point capacity-achieving codes for three pages separately. 2) With NO labeling, for SC decoding, 6 corner points are reduced to 4 points. Moreover, the rate $\left(R_1 = 1, R_2 = 1, R_3 = I(X_3; Y)\right)$ can be achieved, which means that we only need to apply a point-to-point capacity-achieving code to the page $X_3$, and the pages $X_1$ and $X_2$ do not need encoding.

Thanks to the similarity between the data retention model in Table VII and the early stage P/E cycling model in Table VI, for the data retention model, Lemma 18, Theorems 19 and 20 still hold. □

**Example 5** For the early stage P/E cycling model with channel transition matrix $p_{TLC}^E(y|v)$ in Table VI, let $\varepsilon_i = 0.9$, $i = 0, 1, \ldots, 6$. The uniform rate regions of TIN and SC decodings under Gray and NO labelings are plotted in Fig. 8. It is shown that $\mathscr{R}_{NO}^{TIN} \subset \mathscr{R}_G^{TIN} = \mathscr{R}_G^{SC} \subset \mathscr{R}_{NO}^{SC}$. For SC decoding under NO labeling, the corner point $\mathbb{P}_{NO}^5 = \mathbb{P}_{NO}^6 = (R_1 = 1, R_2 = 1, R_3 = 0.5878)$ is achieved. For this particular point, encoding is only needed for the page $X_3$.

Next, we study the structure and property of all labelings. There exist a total of $8! = 40320$ labelings. Similar to MLC flash memory, we treat a *labeling* $\sigma$ as a *permutation* $\pi$ in the *symmetric group* $\mathcal{S}_8$ with group operation $*$. We write a permutation $\pi$ in $\mathcal{S}_8$ as a vector $\pi = (w_0, w_1, w_2, w_3, w_4, w_5, w_6, w_7)$ (where the $w_i$, $i = 0, 1, \ldots, 7$, represent the full set of possible 3-tuples) to represent that $\pi(111) = w_0$, $\pi(110) = w_1$, $\pi(101) = w_2$, $\pi(100) = w_3$, $\pi(011) = w_4$, $\pi(010) = w_5$, $\pi(001) = w_6$, and $\pi(000) = w_7$.

In the following, we will write *decimal* labeling to represent *binary* labeling for simplicity. For example, the binary labeling (111, 110, 101, 100, 011, 010, 001, 000) will be written as (7, 6, 5, 4, 3, 2, 1, 0). By utilizing the algebraic structure of the symmetric group $\mathcal{S}_8$, we have following two lemmas on the uniform rate region and sum rate, valid for any channel model characterized by an *arbitrary* channel transition matrix $p_{TLC}(y|v)$ for $y \in \mathcal{Y}_{TLC}$ and $v \in \mathcal{V}_{TLC}$. We omit their proofs, since they are similar to the MLC case.

**Lemma 21.** *In the symmetric group $\mathcal{S}_8$, $D_0 = \{(7, 6, 5, 4, 3, 2, 1, 0), (6, 7, 4, 5, 2, 3, 0, 1), (5, 4, 7, 6, 1, 0, 3, 2), (4, 5, 6, 7, 0, 1, 2, 3), (3, 2, 1, 0, 7, 6, 5, 4), (2, 3, 0, 1, 6, 7, 4, 5), (1, 0, 3, 2, 5, 4, 7, 6), (0, 1, 2, 3, 4, 5, 6, 7)\}$ forms an abelian subgroup which includes $\sigma_{NO} = (7, 6, 5, 4, 3, 2, 1, 0)$. With an arbitrary channel transition matrix $p_{TLC}(y|v)$, all labelings in $D_0$ give the same uniform rate region for a given decoding scheme (TIN or SC decoding).*

**Lemma 22.** *In the symmetric group $\mathcal{S}_8$, $\hat{D}_0 = \{D_0 \cup D_0*(7, 6, 3, 2, 5, 4, 1, 0) \cup D_0*(7, 3, 5, 1, 6, 2, 4, 0) \cup D_0*(7, 5, 6, 4, 3, 1, 2, 0) \cup D_0*(7, 3, 6, 2, 5, 1, 4, 0) \cup D_0*(7, 5, 3, 1, 6, 4, 2, 0)\}$ forms an abelian subgroup which includes $\sigma_{NO} = (7, 6, 5, 4, 3, 2, 1, 0)$. With an arbitrary channel transition matrix $p_{TLC}(y|v)$, all labelings in $\hat{D}_0$ give the same sum rate for a given decoding scheme (TIN or SC decoding).*

**Remark 5** With subgroup $D_0$ in Lemma 21, we can partition $\mathcal{S}_8$ into $D_0$ and its 5039 cosets, each of size 8. All labelings in each coset give the same uniform rate region for a given decoding scheme (TIN or SC decoding). With subgroup $\hat{D}_0$ in Lemma 22, the group $\mathcal{S}_8$ can be partitioned into $\hat{D}_0$ and its 839 cosets, each of size 48. All labelings in each coset give the same sum rate for TIN decoding. Note that for SC decoding, all labelings in $\mathcal{S}_8$ give the same sum rate, due to the uniform input distributions for $X_1$, $X_2$, and $X_3$.

Thus, for an arbitrary channel transition matrix $p_{TLC}(y|v)$, for TIN decoding, among all the labelings in $\mathcal{S}_8$, the 48 labelings in the coset $\hat{D}_0 * (7, 6, 4, 5, 1, 0, 2, 3)$, which includes Gray labeling, give the largest sum rate, which equals that of SC decoding. However, for some particular channel transition matrix, for TIN decoding, more than 48 labelings may give the largest sum rate. For example, for the channel transition matrix $p_{TLC}^E(y|v)$ in Table VI, for TIN decoding, among all the labelings in $\mathcal{S}_8$, we find the 144 labelings in the set $\tilde{D} = \{\hat{D}_0 * (7, 6, 4, 5, 1, 0, 2, 3) \cup \hat{D}_0 * (7, 6, 4, 0, 2, 3, 1, 5) \cup \hat{D}_0 * (7, 6, 2, 3, 1, 5, 4, 0)\}$ give the largest sum rate. $\qquad\square$

Finally, we discuss the uniform rate region if we are allowed to use multiple labelings together for each codeword instead of one labeling in the TLC flash memory. Define $\mathscr{R}_{\mathcal{S}_8}^{TIN} = Conv\left(\bigcup_{\sigma \in \mathcal{S}_8} \mathscr{R}_\sigma^{TIN}\right)$, the convex hull of uniform rate regions of all 40320 labelings for TIN decoding. Define $\mathscr{R}_{\mathcal{S}_8}^{SC} = Conv\left(\bigcup_{\sigma \in \mathcal{S}_8} \mathscr{R}_\sigma^{SC}\right)$, the convex hull of uniform rate regions of all 40320 labelings for SC decoding. Through *time sharing* of different labelings, for TIN decoding, any point $(R_1, R_2, R_3) \in \mathscr{R}_{\mathcal{S}_8}^{TIN}$ can be achieved. For SC decoding, any point $(R_1, R_2, R_3) \in \mathscr{R}_{\mathcal{S}_8}^{SC}$ can be achieved. Similar to Theorem 13, for the early stage P/E cycling model with channel transition matrix $p_{TLC}^E(y|v)$ in Table VI, the region $\mathscr{R}_{\mathcal{S}_8}^{SC}$ can be determined explicitly. We give the following theorem without proof which is similar to that of Theorem 13.

**Theorem 23.** *For the early stage P/E cycling model, the rate region $\mathscr{R}_{\mathcal{S}_8}^{SC}$ is the set of all pairs $(R_1, R_2, R_3)$ such that:*

1) $R_1 \leqslant 1, R_2 \leqslant 1, R_3 \leqslant 1$;
2) $R_1 + R_2 \leqslant I(X_1, X_2, X_3; Y) - 1, R_1 + R_3 \leqslant I(X_1, X_2, X_3; Y) - 1,$
   $R_2 + R_3 \leqslant I(X_1, X_2, X_3; Y) - 1$;
3) $R_1 + R_2 + R_3 \leqslant I(X_1, X_2, X_3; Y)$.

## VI. Conclusion

We analyzed different decoding schemes for flash memories from a multi-user perspective. In MLC flash memory, both TIN and SC decoding schemes outperform the current default decoding scheme in terms of either rate region or sum rate. For the P/E cycling model, for TIN decoding, 8 labelings which include Gray labeling give the largest sum rate among all 24 labelings. The sum rate of TIN decoding under Gray labeling equals that of SC decoding at the early stage of P/E cycling, and is smaller than but close to that of SC decoding at the late stage of P/E cycling. It was also shown additional read thresholds can effectively enhance the rate region and sum rate. For TLC flash memory, the structure and property of all 40320 labelings were investigated. We studied the early stage P/E cycling model which is similar to the data retention model. For TIN decoding, 144 labelings including Gray labeling give the largest sum rate among all 40320 labelings. Moreover, surprisingly, for SC decoding under NO labeling, the rate $(R_1 = 1, R_2 = 1, R_3)$ can be achieved, implying a simple ECC solution where the first two of the three pages do not need encoding.

## References

[1] E. Arikan, "Channel polarization: A method for constructing capacity-achieving codes for symmetric binary-input memoryless channels," *IEEE Transactions on Information Theory*, vol. 55, no. 7, pp. 3051–3073, July 2009.

[2] M. Asadi, X. Huang, A. Kavcic, and N. P. Santhanam, "Optimal detector for multilevel NAND flash memory channels with intercell interference," *IEEE Journal on Selected Areas in Communications*, vol. 32, no. 5, pp. 825–835, May 2014.

[3] R. Bez, E. Camerlenghi, A. Modelli, and A. Visconti, "Introduction to flash memory," *Proceedings of the IEEE*, vol. 91, no. 4, pp. 489–502, April 2003.

[4] Y. Cai, E. F. Haratsch, O. Mutlu, and K. Mai, "Error patterns in MLC NAND flash memory: Measurement, characterization, and analysis," in *Proc. IEEE Design, Automation & Test in Europe Conference & Exhibition (DATE)*, Dresden, March 2012, pp. 521–526.

[5] ——, "Threshold voltage distribution in MLC NAND flash memory: Characterization, analysis, and modeling," in *Proc. IEEE Design, Automation & Test in Europe Conference & Exhibition (DATE)*, Grenoble, France, March 2013, pp. 1285–1290.

[6] T. M. Cover and J. A. Thomas, *Elements of Information Theory*. Hoboken, New Jersey: John Wiley & Sons, 2012.

[7] G. Dong, S. Li, and T. Zhang, "Using data postcompensation and predistortion to tolerate cell-to-cell interference in MLC NAND flash memory," *IEEE Transactions on Circuits and Systems I: Regular Papers*, vol. 57, no. 10, pp. 2718–2728, Oct. 2010.

[8] G. Dong, N. Xie, and T. Zhang, "On the use of soft-decision error-correction codes in NAND flash memory," *IEEE Transactions on Circuits and Systems I: Regular Papers*, vol. 58, no. 2, pp. 429–439, Feb. 2011.

[9] A. El Gamal and Y.-H. Kim, *Network Information Theory*. Cambridge, UK: Cambridge University Press, 2011.

[10] P. Huang, P. H. Siegel, and E. Yaakobi, "Performance of flash memories with different binary labelings: A multi-user perspective," in *Proc. IEEE International Symposium on Information Theory (ISIT)*, Barcelona, Spain, July 2016.

[11] X. Huang, A. Kavcic, X. Ma, G. Dong, and T. Zhang, "Optimization of achievable information rates and number of levels in multilevel flash memories," in *Proc. the Twelfth International Conference on Networks (ICN)*, Seville, Spain, Jan. 2013, pp. 125–131.

[12] Q. Li, A. Jiang, and E. F. Haratsch, "Noise modeling and capacity analysis for NAND flash memories," in *Proc. IEEE International Symposium on Information Theory (ISIT)*, Honolulu, HI, June 2014, pp. 2262–2266.

[13] J. Moon, J. No, S. Lee, S. Kim, and J. Yang, "Statistical analysis of flash memory read data," in *Proc. IEEE Global Telecommunications Conference (GLOBECOM)*, Houston, TX, Dec. 2011, pp. 1–6.

[14] T. Parnell, N. Papandreou, T. Mittelholzer, and H. Pozidis, "Modelling of the threshold voltage distributions of sub-20nm NAND flash memory," in *Proc. IEEE Global Communications Conference (GLOBECOM)*, Austin, TX, Dec. 2014, pp. 2351–2356.

[15] A. Paszkiewicz and T. Sobieszek, "Additive entropies of partitions," *arXiv preprint arXiv:1202.4591*, 2012.

[16] B. Peleato, R. Agarwal, J. M. Cioffi, M. Qin, and P. H. Siegel, "Adaptive read thresholds for NAND flash," *IEEE Transactions on Communications*, vol. 63, no. 9, pp. 3069–3081, Sept. 2015.

[17] W. J. Reed, "The normal-Laplace distribution and its relatives," in *Advances in distribution theory, order statistics, and inference*. Springer, 2006, pp. 61–74.

[18] C. Schoeny, F. Sala, and L. Dolecek, "Analysis and coding schemes for the flash normal-Laplace mixture channel," in *Proc. IEEE International Symposium on Information Theory (ISIT)*, Hongkong, June 2015, pp. 2101–2105.

[19] V. Taranalli, H. Uchikawa, and P. H. Siegel, "Error analysis and inter-cell interference mitigation in multi-level cell flash memories," in *Proc. IEEE International Conference on Communications (ICC)*, London, June 2015, pp. 271–276.

[20] J. Wang, T. Courtade, H. Shankar, and R. D. Wesel, "Soft information for LDPC decoding in flash: Mutual-information optimized quantization," in *Proc. IEEE Global Telecommunications Conference (GLOBECOM)*, Houston, TX, Dec. 2011, pp. 1–6.

[21] J. Wang, K. Vakilinia, T.-Y. Chen, T. Courtade, G. Dong, T. Zhang, H. Shankar, and R. Wesel, "Enhanced precision through multiple reads for LDPC decoding in flash memories," *IEEE Journal on Selected Areas in Communications*, vol. 32, no. 5, pp. 880–891, May 2014.